\documentclass[preprint]{ptephy_v1}

\preprintnumber{XXXX-XXXX} 
\usepackage{hyperref}
\usepackage{ulem}
\usepackage{color}

\begin{document}

\title{A new boson expansion theory utilizing a norm operator}


\author{Kimikazu Taniguchi}
\affil{Department of Medical Information Science, Suzuka University of Medical Science, Suzuka  1001-1, Japan\email{kimikazu@suzuka-u.ac.jp}}





\begin{abstract}
We propose a new boson expansion method using a norm operator. The small parameter expansion, in which the boson approximation becomes the zeroth-order approximation, requires the double commutation relations between phonon operators that are not closed between the phonon excitation modes adopted as boson excitations. 
This results in an infinite expansion regardless of whether the type of the boson expansion is Hermitian or non-Hermitian.
The small parameter expansion does not hold when the commutation relations are closed. 
The norm operator is expressed as a function of the number operator in the physical subspace, which enables us to obtain substantially a finite boson expansion regardless of the Hermitian or non-Hermitian type. 
We also point out the problems of the conventional boson expansion methods.
The normal-ordered linked-cluster expansion theory has failed to refute Marshalek's claim that KT-1 and KT-2 are of chimerical boson expansion. 
The Dyson boson expansion theory does not have exceptional superiority over other types.
Previous studies using the boson expansion methods should be re-examined.
\end{abstract}

\subjectindex{xxxx, xxx}

\maketitle

\section{Introduction}
Microscopic elucidation of the large-amplitude collective motion of atomic nuclei remains one of the important and challenging tasks, and its achievement requires developing a method to overcome small-amplitude-oscillation approximations like the Tamm-Dancoff approximation and the random phase approximation. The boson expansion theory is one of the methods going beyond the small-amplitude-oscillation approximations \cite{KM91}. 

The boson expansion theory was initially formulated by replacing fermion quasi-particle pair operators with the boson polynomials that reproduce their commutation relations \cite{BZ62}.
Later, referring to the preceding work \cite{Usi60}, the boson expansion theory has been given as a mapping theory by utilizing a one-to-one correspondence between the basis vector in fermion space and the completely antisymmetric state vector in boson space \cite{MYT64}. The Holstein-Primakoff and the Dyson Boson expansions have been formulated also in the same way \cite {JDF71}.
These formulations target all pair operator excitations in fermion space.

Practical use, however, has not adopted all excitation modes but the collective and, in case of need, some non-collective excitation modes of Tamm-Dancoff type phonons.
Initially, there were two methods:
One is to construct the mapping operator by the phonons with only the crucial excitation modes \cite{LH75, HJJ76}, and the other is the method to pick up only these phonons and seek boson expansions that reproduce their commutation relations \cite {KT76}.  
Thus formulated boson expansion methods were used for elucidating the large-amplitude collective motions such as the shape transitions of nuclei in the transitional region \cite{KT76, TTT87, SK91}. 
The boson expansion method called KT-2 \cite {KT76} formulated in the latter method was, however, claimed to result in incorrect boson expansions \cite{Ma80a, Ma80b} and reformulated into so-called KT-3 \cite{KT83} according to the former method. 
The Dyson boson expansion theory (DBET), finite expansion of the non-Hermitian type, has been also formulated by the former method \cite {Ta01}.

Seeming to establish the formulation and achieve certain results, the problems remain with the boson expansion methods.

One is on the approximate treatment of the algebra of the Tamm-Dancoff type phonons.
The double commutators among the Tamm-Dancoff type phonons generally do not close within partially selected excitation modes.
Until now without exception, the boson expansion methods with restriction of the phonon excitation modes have used approximations that neglect the modes not selected for the boson excitation modes.
The normal-ordered linked-cluster expansion theory (NOLCEXPT) \cite{KT83, SK88} neglects these in the inverse of the norm matrices of the multi-phonon state vectors to obtain its boson expansions and finally abandons all the still-remaining modes.
DBET truncates the unselected phonon operators by adopting the approximation named phonon-truncation approximation \cite{TTT87},
which is also called {\it closed-algebra approximation} \cite{Ta01}.
Each of the approximations above is essential for NOLCEXPT and DBET. NOLCEXPT adopts it for obtaining the same expansions as KT-2, and DBET to obtain the finite expansions.
These approximations all bring to make the double commutators among the selected phonon operators closed.
It is claimed that the validity of this approximation has been verified for the specific nuclei, and it is also shown that the norm of the multi-phonon state vector obtained under this approximation rapidly approaches 0 as the number of phonon excitations increases, which brings rapid convergence of the boson expansions \cite{HJJ76, MTS81}. 
Such behavior of the norm is due to the effect of the Pauli exclusion principle \cite{MTS81}. Its rapid decrease means that the effect is strong.
On the other hand, NOLCEXPT claims that its boson expansion is of a small parameter expansion with good convergence.
Therefore, in the fermion subspace spanned by the multi-phonon state vectors with selected excitation modes, the effect of the Pauli exclusion principle should be weak.
If this is correct, then the norms of the multi-phonon state vectors would not approach zero rapidly as the number of phonon excitations increases. 
We should investigate the cause of contradictory conclusions.

The other is about the phonon excitation number. Until now, for the multi-phonon state vectors, used as the basis vectors of the fermion subspace to be mapped, the sorts of the excitation modes have been limited, while the number of phonon excitations has not \cite{KT83, SK88, Ta01}. Without restricting the phonon excitation number, the eigenvalues of the norm matrices of the multi-phonon state vectors become zero when the number of excitations becomes large enough even with restricting the sorts of excitation modes. 
Nevertheless, NOLCEXPT is formulated assuming that zero eigenvalues do not appear regardless of the number of phonon excitations\cite{KT83, SK88}. There is, however, no clear explanation for the validity of this assumption.

We have proposed a boson-fermion expansion theory (BFEXP) \cite{TM90, TM91} as an alternative to the boson expansion theory.
The boson expansion theory treats all the adopted phonon excitation modes as bosons, while BFEXP, in the zeroth order approximation, represents only the phonons with collective excitation modes as bosons and those with non-collective remains as original phonons. 
We can derive the boson expansions from this method by extending the boson part up to the non-collective modes necessary and depressing the fermion excitations. 
Since the formulation of BFEXP has not used the approximation for commutation relations among the phonon operators, 
it would be worthwhile to formulate a new boson expansion method without the approximation for the commutation relations among phonon operators and compare its boson expansions with those derived from BFEXP.

In this article, we propose a new boson expansion theory, naming it the norm operator method (NOM), which enables us to handle both Hermitian and non-Hermitian types, the case with or without limiting the phonon excitation modes and the number of excitations, and the contribution of the phonon excitation modes which are neglected in the conventional boson expansion methods.

In section \ref{fsbs}, we deal with the Tam-Dancoff type phonons, the multi-phonon state vectors, and the ideal boson state vectors.

In section \ref{bmap}, we give a mapping utilizing a norm operator.
As specific examples, we deal with the case of mapping all modes of phonon excitations with and without the restriction of the phonon excitation number and the restricted case where the maximum number of phonon excitations is one.

Section \ref{bexp} deals with the boson expansions. First, we confirm the conditions for using the ideal boson state vectors and then give the formulae used in the boson expansions. Next, we provide the conditions that boson expansions become of a small parameter expansion, offer an order estimation method for the expansion terms, perform the boson expansions,
show that all types of mapping of the small parameter expansion give infinite boson expansions,
and provide the boson expansions of the phonon operators and the scattering operators up to terms that have not been obtained so far.
We also deal with non-small parameter boson expansions, where we obtain DBET and the boson expansions that are finite and Hermite.
Finally, we point out and stress the essential role of the norm operator in the boson expansion method.

In section \ref{comp}, we take up the conventional methods and point out their problems.

In section \ref{appl}, we comment on its application to the collective motions of nuclei.

Section \ref{sum} is a summary.

\section{Fermion space and boson space}
\label{fsbs}
\subsection{Tamm-Dancoff type phonon operators, scattering operators, and their commutation relations
}
We introduce pair operators,
\begin{subequations}
\label{eq:phononop}
\begin{equation}
\label{eq:phononopc}
X_\mu^\dagger
=\displaystyle\sum_{\alpha<\beta}\psi_\mu(\alpha\beta)a_\alpha^\dagger
a_\beta^\dagger,
\end{equation}
\begin{equation}
\label{eq:phononopa}
X_\mu=\displaystyle\sum_{\alpha<\beta}\psi_\mu(\alpha\beta)a_\beta
a_\alpha,
\end{equation}
\end{subequations}
\begin{subequations}
\label{eq:scop12}
\begin{equation}
\label{eq:scop}
B_q=\sum_{\alpha\beta}\varphi_q(\alpha\beta)a_\beta^\dagger
a_\alpha,
\end{equation}
\begin{equation}
\label{eq:scop2}
B_{\bar q=}B_q^\dagger.
\end{equation}
\end{subequations}
Here, $a_\alpha^\dagger$ and $a_\alpha$ are quasi-particle creation and annihilation operators in a single-particle state $\alpha$.
The coefficients satisfy the following relations:
\begin{subequations}
\label{eq:TDrel}
\begin{equation}
\label{eq:anti}
\psi_\mu(\beta\alpha)=-\psi_\mu(\alpha\beta)
\end{equation}
\begin{equation}
\label{eq:orthonormal}
\sum_{\alpha<\beta}\psi_\mu(\alpha\beta)\psi_{\mu'}(\alpha\beta)=\delta_{\mu,
\mu'},
\end{equation}
\begin{equation}
\label{eq:complete}
\sum_{\mu}\psi_\mu(\alpha\beta)\psi_{\mu}(\alpha'\beta')=\delta_{\alpha,
\alpha'}\delta_{\beta, \beta'}-\delta_{\alpha,
\beta'}\delta_{\beta, \alpha'},
\end{equation}
\end{subequations}
\begin{subequations}
\label{eq:coeffscop}
\begin{equation}
\label{eq:barq}
\varphi_{\bar q}(\alpha\beta)=\varphi_q(\beta\alpha).
\end{equation}
\begin{equation}
\label{eq:coeffscop1}
\displaystyle\sum_{\alpha\beta}\varphi_q(\alpha\beta)\varphi_{q'}(\alpha\beta)=\delta_{q,q'},
\end{equation}
\begin{equation}
\label{eq:coeffscop2}
\displaystyle\sum_{q}\varphi_q(\alpha\beta)\varphi_{q}(\alpha'\beta')=\delta_{\alpha,\alpha'}\delta_{\beta,
\beta'}.
\end{equation}
\end{subequations}
These are the most common orthogonal transformations of the quasi-particle pairs $a_\alpha^\dagger a_\beta^\dagger$, $a_\beta a_\alpha$ and $a_\beta^\dagger a_\alpha$. These are used to couple the angular momenta of quasi-particles to those of the quasi-particle pairs. Some of $X_\mu$ and $X_\mu^\dagger$ are composed by the further superposition of such pair operators to reflect the dynamics into the selected phonons. Tamm-Dancoff approximation or a similar approximation is usually applied to them for identifying collective excitation modes and non-collective ones. Hereafter, $X_\mu^\dagger$ and $X_\mu$ are called phonon creation and annihilation operators, and $B_q$ is called a scattering operator.

The phonon and scattering operators satisfy the following commutation relations:
\begin{subequations}
\label{eq:algebra}
\begin{equation}
\label{eq:algebra1}
[ X_\mu, X_{\mu'}^\dagger ]=\delta_{\mu,
\mu'}-\sum_q\Gamma^{\mu\mu'}_qB_q,
\end{equation}

\begin{equation}
\label{eq:algebra2}
[ B_q, X_\mu^\dagger
]=\sum_{\mu'}\Gamma^{\mu\mu'}_qX_{\mu'}^\dagger,
\end{equation}

\begin{equation}
\label{eq:algebra3}
[ X_\mu, B_q ]=\sum_{\mu'}\Gamma^{\mu'\mu}_qX_{\mu'},
\end{equation}
\end{subequations}
where the definition of $\Gamma^{\mu\mu'}_q$ is as follows:
\begin{equation}
\label{eq:Gamma}
\Gamma^{\mu\mu'}_q=\sum_{\alpha\beta}\varphi_q(\alpha\beta)\Gamma^{\mu\mu'}_{\alpha\beta},\quad
\Gamma^{\mu\mu'}_{\alpha\beta}=\sum_\gamma\psi_\mu(\alpha\gamma)\psi_{\mu'}(\beta\gamma).
\end{equation}
The following relation holds:
\begin{equation}
\label{eq:qbargam}
\Gamma_{\bar q}^{\mu_1 \mu_2}=\Gamma_q^{\mu_2 \mu_1}.
\end{equation}

From Eqs. (\ref{eq:algebra1}) and (\ref{eq:algebra2}), we obtain
\begin{equation}
\label{eq:doublecom}
[ [X_{\mu_1}, X_{\mu_2}^\dagger], X_{\mu_3}^\dagger] = -\sum_{\mu'}Y(\mu_1, \mu_2, \mu_3, \mu')X_{\mu'}^\dagger,
\end{equation}
where the definition of $Y(\mu_1\mu_2\mu_3\mu_4)$ is 
\begin{equation}
\label{eq:Y}
Y(\mu_1\mu_2\mu_3\mu_4)=\sum_q\Gamma_q^{\mu_1\mu_2}\Gamma_q^{\mu_3\mu_4}
=\sum_{\alpha\beta}\Gamma_{\alpha\beta}^{\mu_1\mu_2}\Gamma_{\alpha\beta}^{\mu_3\mu_4}.
\end{equation}
The following relation holds:
\begin{equation}
\label{eq:Ysym}
\begin{array}{lll}
Y(\mu_1\mu'_1\mu'_2\mu_2)
&=&Y(\mu_2\mu'_1\mu'_2\mu_1)
\\
&=&Y(\mu_1\mu'_2\mu'_1\mu_2)
\\
&=&Y(\mu'_1\mu_1\mu_2\mu'_2).
\\
\end{array}
\end{equation}

\subsection{Multi-phonon and multi-boson state vectors}
We divide the phonon excitation modes $\{\mu\}$ into two groups, $\{t\}$ and $\{\bar t\}$,and prepare the multi-phonon state vectors, 
\begin{equation}
\label{eq:mulphst}
\vert N; t\rangle\rangle=\vert t_1, t_2, \cdots ,
t_N\rangle\rangle=X_{t_1}^\dagger X_{t_2}^\dagger \cdots
X_{t_N}^\dagger \vert 0\rangle\quad (0\leq N\leq N_{max}).
\end{equation}
$\{t\}$ usually consists of collective modes and some non-collective modes if necessary, selected by the small amplitude approximation. We treat not only these cases but also the case where all modes are adopted, that is, $\{t\}=\{\mu\}$.

Next we introduce boson creation and annihilation operators, $b_t^\dagger$ and $b_{t'}$, having the same indices as those of the multi-phonons, $X_t^\dagger$ and $X_{t'}$:
\begin{equation}
\label{eq:bcom}
[ b_t, b_{t'}^\dagger ] = \delta_{t, t'}.
\end{equation}
The multi-boson states,
\begin{equation}
\label{eq:multib}
\vert N; t))=\vert t_1, t_2, \cdots , t_N))=b_{t_1}^\dagger
b_{t_2}^\dagger \cdots b_{t_N}^\dagger \vert 0),
\end{equation}
are orthogonal to one another, and are normalized by their norms,
\begin{equation}
\label{eq:normb}
\mathcal{N}_B(N; t)=((N: t\vert N; t)),
\end{equation}
such as
\begin{equation}
\label{eq:bbasis}
\vert N; t)=\vert t_1, t_2, \cdots , t_N)=\mathcal{N}_B(N; t)^{-1/2}\vert N; t)).
\end{equation}
They are so-called ideal boson state vectors.

\section{Boson mapping}
\label{bmap}
This section deals with boson mapping. We introduce a norm operator and construct a mapping operator that can handle both Hermitian and non-Hermitian types, both with and without limiting the types and number of phonon excitation modes.

The norm operator is defined as
\begin{subequations}
\begin{equation}
\label{eq:normop}
\hat Z=\sum_{N=0}^{N_{max}}\hat Z(N),
\end{equation}
\begin{equation}
\label{eq:normop2}
\begin{array}{lll}
\hat{Z}(N)&=&\displaystyle\sum_{t t'}\vert N, t)\langle N;
t\vert N; t'\rangle ( N; t' \vert
\\
&=&
\displaystyle\sum_{t_1\leq\cdots \leq t_N}\sum_{t'_1\leq\cdots
\leq t'_N}\vert t_1 \cdots t_N)\langle t_1\cdots t_N\vert
t'_1\cdots t_N\rangle (t'_1\cdots t'_N \vert,
\end{array}
\end{equation}
\end{subequations}
where 
\begin{equation}
\label{eq:bnmulphst}
\vert N; t\rangle=\mathcal{N}_B(N; t)^{-1/2}\vert N;
t\rangle\rangle.
\end{equation}

This norm operator is a modified one of the previously introduced \cite{KT83} by adding the restriction $N_{max}$, which allows us to constrain the number of phonon excitations and corresponding boson excitations.
$\hat Z(N)$ satisfies the eigenequation,
\begin{equation}
\label{eq:eigeneqnrop1}
\hat Z(N)\vert N; a)=z_a(N)\vert N; a),
\end{equation}
where $\vert N; a)$ is a normalized eigenvector and $z_a(N)$ is an eigenvalue.
The eigenvalues $z_a(N)$ become positive or zero and $a_0$ represents $z_{a_0}(N)=0$.
Using these,
we obtain the spectral decomposition of $\hat Z(N)$ as
\begin{equation}
\label{eq:nropspec}
\hat{Z}(N) = \sum_{a\neq a_0}\vert N; a)z_a(N)(N; a\vert.
\end{equation}
Functions of $\hat Z(N)$ are defined by
\begin{equation}
f(\hat Z(N))=\sum_{a\neq a_0}\vert N; a)f(z_a(N))(N; a\vert,
\end{equation}
and we obtain
\begin{equation}
f(\hat Z)=\sum_{N=0}^{N_{max}}f(\hat Z(N)).
\end{equation}
Introducing $u_a^t(N)=(N; t\vert N;a)$, then Eq. (\ref{eq:eigeneqnrop1}) becomes the eigenequation of the multi-phonon norm matrix,
\begin{equation}
\label{eq:eigeneqnormmtx}
\sum_{t'}\langle N; t\vert N; t'\rangle u_a^{t'}(N)=z_a(N)u_a^t(N),
\end{equation}
The eigenvectors are orthonomalized as
\begin{subequations}
\label{eq:eigenvcon}
\begin{equation}
\label{eq:eigenvcon1}
\displaystyle\sum_tu_a^t(N)u_{a'}^t(N)=\delta_{a, a'},
\end{equation}
and satisfy the completeness relations
\begin{equation}
\label{eq:eigenvcon2}
\displaystyle\sum_au_a^t(N)u_a^{t'}(N)=\delta_{t, t'}.
\end{equation}
\end{subequations}

Using the norm operator $\hat Z$, we introduce the mapping operator $U_\xi$ as
\begin{equation}
\label{eq:bmopzubar}
U_\xi=\hat Z^{\xi-\frac 12}\widetilde U,
\end{equation}
where $\widetilde U$ is a mapping operator whose definition is as follows:
\begin{subequations}
\label{eq:tildempop}
\begin{equation}
\label{eq:tildempop1}
\widetilde {U}=\sum_{N=0}^{N_{max}}{\widetilde U}(N),
\end{equation}
\begin{equation}
\label{eq:tildempop2}
\begin{array}{lll}
{\widetilde U}(N)&=&\displaystyle\sum_{t}\vert N; t)\langle N; t\vert
\\
&=&
\displaystyle\sum_{t_1\leq t_2\leq\cdots\leq t_N}\vert t_1 t_2\cdots
t_N)\langle t_1 t_2\cdots t_N\vert,
\end{array}
\end{equation}
\end{subequations}
which satisfies the following relations,
\begin{subequations}
\label{eq:tildeurel}
\begin{equation}
\label{eq:tldeurel1}
\widetilde U{\widetilde U}^\dagger =\hat{Z},
\end{equation}
\begin{equation}
\label{eq:tildeurel2}
{\widetilde U}(N){\widetilde U}(N)^\dagger =\hat{Z}(N).
\end{equation}
\end{subequations}
${\widetilde U}(N)$ is also expressed as
\begin{equation}
{\widetilde U}(N)=\sum_{a\neq a_0}z_a^{\frac 12}\vert N; a)\langle N; a\vert,
\end{equation}
where
\begin{equation}
\label{eq:orthonormbs}
\vert N; a\rangle = z_a^{-\frac 12}(N)\sum_tu_a^t(N)\vert N; t\rangle
\qquad (a\neq a_0).
\end{equation}
$\vert N;a\rangle$ become orthonormalized basis vectors of the fermion subspace spanned by $\vert N; t\rangle$.
Using $\vert N: a\rangle$ and $\vert N: a)$, the mapping operator is expressed as
\begin{equation}
\label{eq:bmop}
U_\xi=\sum_{N=0}^{N_{max}}U_\xi(N);\quad U_\xi(N)=\sum_{a\neq a_0}z_a(N)^\xi\vert N; a)\langle N; a\vert.
\end{equation}
The following relations are satisfied:
\begin{equation}
\label{eq:utftbh}
U_{-\xi}^\dagger U_{\xi}=\hat T_F,\qquad U_{\xi}U_{-\xi}^\dagger =\hat T_B,
\end{equation}
where
\begin{equation}
\label{eq:unitf}
\hat T_F=\sum_{N=0}^{N_{max}}\hat T_F(N);\qquad
\hat T_F(N)=\displaystyle\sum_{a\neq a_0}\vert N; a\rangle\langle N; a\vert,
\end{equation}
\begin{equation}
\hat T_B=\sum_{N=0}^{N_{max}}\hat T_B(N);\quad\hat T_B(N)=\sum_{a\neq a_0}\vert N; a)(N; a\vert.
\end{equation}
In addition, we define the following operators,
\begin{equation}
\label{eq:breve1B}
\breve 1_B=\sum_{N=0}^{N_{max}}\hat 1_B(N);\qquad \hat 1_B(N)=\sum_t\vert N; t)(N;t\vert.
\end{equation}
If $\hat Z(N)$ has even one zero eigenvalue, then $\hat T_B(N)\neq \hat 1_B(N)$ and hence $\hat T_B\neq \breve 1_B$. Otherwise, they match one another.

The state vectors and operators of fermion space are mapped onto those of boson subspace as
\begin{subequations}
\label{eq:ximap}
\begin{equation}
\label{eq:ximap1}
\vert \psi')_{\xi}= U_{\xi}\vert\psi'\rangle,\qquad {}_{-\xi} (\psi\vert =\langle\psi\vert U_{-\xi}^\dagger,
\end{equation}
\begin{equation}
\label{eq:ximap2}
(O_F)_{\xi}=U_{\xi}O_FU_{-\xi}^\dagger.
\end{equation}
\end{subequations}
These satisfy the following relations:
\begin{subequations}
\label{eq:ximappm}
\begin{equation}
\label{eq:ximappm1}
\vert \psi')_{\xi}=\left\{{}_\xi(\psi'\vert\right\}^\dagger,\qquad {}_{-\xi} (\psi\vert =\left\{\vert\psi)_{-\xi}\right\}^\dagger,
\end{equation}
\begin{equation}
\label{eq:ximappm2}
(O_F)_{-\xi}=\left\{(O_F^\dagger)_{\xi}\right\}^\dagger.
\end{equation}
\end{subequations}
The mapping is of the Hermitian type when $\xi=0$ and, in other cases, of the non-Hermitian type.

A one-to-one correspondence exists between the fermion subspace projected by $\hat T_F$ and the boson subspace by $\hat T_B$.
For the state vectors, $\vert\psi\rangle$ and $\vert\psi'\rangle$, which belong to the fermion subspace projected by $\hat T_F$,
\begin{equation}
\label{eq:mteqh}
\begin{array}{lll}
\langle\psi\vert O_F\vert\psi'\rangle
&=&\langle\psi\vert\hat T_F O_F\hat T_F\vert\psi'\rangle
\\
&=&\langle\psi\vert U_{-\xi}^\dagger U_\xi O_FU_{-\xi}^\dagger U\vert\psi'\rangle
\\
&=&{}_{-\xi}(\psi\vert(O_F)_\xi\vert\psi')_\xi,
\end{array}
\end{equation}
that is, the matrix element of the fermion subspace becomes equal to that of the corresponding boson subspace. 
The boson subspace corresponding to the fermion subspace projected by $\hat T_F$ is called the physical subspace, and the boson state vectors belonging to that space are called the physical state vector. The projection operator of the physical subspace is $\hat T_B$.

The relation
\begin{equation}
{}_{\xi}(\psi\vert(O_F)_{-\xi}\vert\psi')_{-\xi}={}_{-\xi}(\psi\vert(O_F)_\xi\vert\psi')_\xi
\end{equation} 
holds, therefore it is sufficient to treat the case $\xi\ge 0$.

The mapping of the product of the fermion operators does not generally result in the product of the mapped fermion operators.
That is 
\begin{equation}
\label{eq:ximapzbar22}
(O_FO'_F)_{\xi}\neq (O_F)_{\xi}(O'_F)_{\xi},
\end{equation}
and therefore, the commutation relations of the fermion operators are mapped as
\begin{equation}
([O_F, O'_F])_\xi=(O_FO'_F)_\xi-(O'_FO_F)_\xi\neq[(O_F)_\xi, (O'_F)_\xi],
\end{equation}
while under the approximation $O_FO'_F\approx O_F\hat T_FO'_F$, 
\begin{equation}
\label{eq:ximapzbar22aprox}
(O_FO'_F)_{\xi}\approx (O_F)_\xi(O'_F)_\xi,
\end{equation}
and 
\begin{equation}
([O_F, O'_F])_\xi\approx [(O_F)_\xi, (O'_F)_\xi]
\end{equation}
hold.
The conventional practical boson expansion methods use this approximation.
If this approximation holds, it is sufficient to map the phonon and scattering operators, otherwise, it becomes necessary to obtain the mapping of the product of these fermion operators.

We denote the mapping of $\widetilde U$ as
\begin{subequations}
\label{eq:tildemap}
\begin{equation}
\label{eq:tildemap1}
\widetilde{\vert \psi)}= \widetilde U\vert\psi\rangle,\qquad\widetilde{(\psi\vert} =\langle\psi\vert {\widetilde U}^\dagger,
\end{equation}
\begin{equation}
\label{eq:tildemap2}
\widetilde{O_F}=\widetilde UO_F{\widetilde U}^\dagger.
\end{equation}
\end{subequations}
The mapping of Eqs. (\ref{eq:ximap}) is expressed as
\begin{subequations}
\label{eq:ximapzbar}
\begin{equation}
\label{eq:ximapzbar1}
\vert \psi')_{\xi}= \hat Z^{\xi-\frac 12}\widetilde{\vert\psi')},\qquad {}_{-\xi} (\psi\vert=\widetilde{(\psi\vert}\hat Z^{-\xi-\frac 12},
\end{equation}
\begin{equation}
\label{eq:ximapzbar2}
(O_F)_{\xi}=\hat Z^{\xi-\frac 12}\widetilde{O_F}\hat Z^{-\xi-\frac 12},
\end{equation}
\end{subequations}
which makes it clear that the different treatment of the norm operator in the mapping operator produces another type of mapping.
The mapping of Eqs. (\ref{eq:ximap}) is also expressed as
\begin{subequations}
\label{eq:nhhr}
\begin{equation}
\label{eq:nhhr1}
\vert \psi')_{\xi}= \hat Z^{\xi}\vert\psi')_0,\qquad {}_{-\xi} (\psi\vert={}_0(\psi\vert\hat Z^{-\xi},
\end{equation}
\begin{equation}
\label{eq:nhhr2}
(O_F)_{\xi}=\hat Z^{\xi}(O_F)_0\hat Z^{-\xi},
\end{equation}
\end{subequations}
The mapping of $\xi=0$ being of the Hermitian type and that of $\xi\neq 0$ being of the non-Hermitian type transform one another by the similarity transformation operator that becomes the power of the norm operator $\hat Z$.

\subsection{The case where all the phonon excitation modes are adopted as the boson excitation modes}
Hereafter, we attach $(A)$ such as $\hat Z^{(A)}$ in the case that we introduce boson operators corresponding to all phonon excitation modes for no confusion.

We start with the following,
\begin{equation}
\begin{array}{c}
\displaystyle\sum_{\mu_1\leq\cdots\leq\mu_N}\vert\mu_1, \cdots, \mu_N\rangle\langle\mu_1, \cdots, \mu_N\vert
=\displaystyle\frac{1}{N!}\sum_{\mu_1\cdots,\mu_N}\vert\mu_1, \cdots, \mu_N\rangle\rangle\langle\langle\mu_1, \cdots, \mu_N\vert
\\
=\displaystyle\frac{1}{2^{2N}N!}
\sum_{\mu_1\cdots,\mu_N}
\sum_{\begin{array}{c}
\alpha_1\beta_1\cdots\alpha_N\beta_N\\
\alpha'_1\beta'_1\cdots\alpha'_N\beta'_N
\end{array}}
\psi_{\mu_1}(\alpha_1\beta_1)\psi_{\mu_1}(\alpha'_1\beta'_1)\cdots\psi_{\mu_N}(\alpha_N\beta_N)\psi_{\mu_N}(\alpha'_N\beta'_N)
\\
\vert\alpha_1\beta_1\cdots\alpha_N\beta_N\rangle\langle\alpha_1\beta_1\cdots\alpha_N\beta_N\vert
\\
=\displaystyle\frac{1}{2^NN!}\sum_{\alpha_1\beta_1\cdots\alpha_N\beta_N}\vert\alpha_1\beta_1\cdots\alpha_N\beta_N\rangle\langle\alpha_1\beta_1\cdots\alpha_N\beta_N\vert
\\
=\displaystyle\frac{(2N)!}{2^NN!}\sum_{\alpha_1\beta_1\leq\cdots\leq\alpha_N\beta_N}\vert\alpha_1\beta_1\cdots\alpha_N\beta_N\rangle\langle\alpha_1\beta_1\cdots\alpha_N\beta_N\vert,
\end{array}
\end{equation}
where
\begin{equation}
\vert\alpha_1\beta_1\cdot\alpha_N\beta_N\rangle=a_{\alpha_1}^\dagger a_{\beta_1}^\dagger\cdots a_{\alpha_N}^\dagger a_{\beta_N}^\dagger\vert 0\rangle,
\end{equation}
and we use that the function $f(t_1, \cdots, t_N)$, which is completely symmetric for the argument, satisfies the following \cite{KKS91}:\begin{equation}
\sum_{t_1\leq\cdots\leq t_N}f(t_1, \cdots , t_N)=\sum_{t_1, \cdots , t_N}\frac{\mathcal{N}_B(t_1, \cdots , t_N)}{N!}f(t_1, \cdots ,t_N).
\end{equation}
From above, we obtain the following relation,
\begin{equation}
\sum_{\mu_1\leq\cdots\leq\mu_N}\vert\mu_1, \cdots, \mu_N\rangle\langle\mu_1, \cdots, \mu_N\vert=(2N-1)!!\hat 1_F^{(A)}(N),
\end{equation}
where
\begin{equation}
\hat 1_F^{(A)}(N)=\sum_{\alpha_1\beta_1\leq\cdots\leq\alpha_N\beta_N}\vert\alpha_1\beta_1\cdots\alpha_N\beta_N\rangle\langle\alpha_1\beta_1\cdots\alpha_N\beta_N\vert;
\quad (2N-1)!!=\frac{(2N)!}{2^NN!}.
\end{equation}
Let $\mathbf{Z}^{(A)}(N)$ be a matrix composed of the matrix element $\langle\mu'_1,\cdots\mu'_N\vert\mu_1\cdots\mu_N\rangle$, we obtain, from this relation,
\begin{equation}
\label{eq:znarelmt}
\mathbf{Z}^{(A)}(N)^2=(2N-1)!!\mathbf{Z}^{(A)}(N),
\end{equation}
which indicates that the eigenvalues of this matrix are $(2N-1)!!$ or $0$. Zero eigenvalues appear even at $N=2$ \cite{KT83}, and so do in the case $N\ge 2$. From this relation we obtain
\begin{equation}
\label{eq:znarel}
\hat Z^{(A)}(N)^2=(2N-1)!!\hat Z^{(A)}(N),
\end{equation}
and
\begin{equation}
\label{eq:zacondi}
\left (\hat Z^{(A)}\right )^2=(2\hat N_B^{(A)}-1)!!Z^{(A)};\quad\hat N_B^{(A)}=\sum_\mu b_\mu^\dagger b_\mu.
\end{equation}
The case N=2 in Eq. (\ref{eq:znarelmt}) is equivalent to the following relations \cite{KT76},
\begin{equation}
\label{eq:yrel}
\sum_{\mu\mu'}Y(\mu'_1\mu\mu'\mu'_2)Y(\mu_1\mu\mu'\mu_2)=4((\mu'_1\mu'_2\vert\mu_1\mu_2))-2Y(\mu'_1\mu_1\mu_2\mu'_2).
\end{equation}

We introduce the following operators,
\begin{equation}
b_{\alpha\beta}=\sum_\mu\psi_\mu(\alpha\beta)b_\mu, \quad b_{\alpha\beta}^\dagger=\sum_\mu\psi_\mu(\alpha\beta)b_\mu^\dagger,
\end{equation}
which satisfies the commutation relations,
\begin{equation}
[b_{\alpha'\beta'}, b_{\alpha\beta}^\dagger]=\delta_{\alpha'\alpha}\delta_{\beta'\beta}-\delta_{\alpha'\beta}\delta_{\beta'\alpha}.
\end{equation}
Using these operators, $\widetilde{U}^{(A)}(N)=\displaystyle\sum_\mu\vert N; \mu)\langle N; \mu\vert$ are rewritten as
\begin{equation}
\widetilde{U}^{(A)}(N)=\sqrt{(2N-1)!!}\sum_{\alpha_1<\beta_1<\cdots<\alpha_N<\beta_N}\vert\alpha_1\beta_1\cdots\alpha_N\beta_N)_M\ \langle\alpha_1\beta_1\cdots\alpha_N\beta_N\vert,
\end{equation}
where
\begin{equation}
\vert\alpha_1\beta_1\cdots\alpha_N\beta_N)_M=\frac{1}{\sqrt{(2N-1)!!}}{\sum_{P}}'(-)^Pb_{\alpha_1\beta_1}^\dagger\cdots b_{\alpha_N\beta_N}^\dagger\vert 0),
\end{equation}
and $\displaystyle{\sum_{P}}'$ means the summation so that the states on the left side become totally antisymmetric \cite{MYT64}.
From these, we obtain
\begin{equation}
\begin{array}{lll}
\hat Z^{(A)}(N)&=&\widetilde{U}^{(A)(N)}\widetilde{U}^{(A)}(N)^\dagger
\\
&=&\displaystyle (2N-1)!!\displaystyle\sum_{\alpha_1<\beta_1<\cdots\alpha_N\beta_N}\vert\alpha_1\beta_1\cdots\alpha_N\beta_N)_M{}_M(\alpha_1\beta_1\cdots\alpha_N\beta_N\vert,
\end{array}
\end{equation}
which is the spectral decomposition of $\hat Z^{(A)}(N)$ and indicates that the eigenvectors of the eigenvalue $(2N-1)!!$ are $\vert\alpha_1\beta_1\cdots\alpha_N\beta_N)_M$.
We also obtain
\begin{equation}
\begin{array}{lll}
\hat T_F^{(A)}&=&\displaystyle\sum_{N=0}^{N_{max}}\hat T_F^{(A)}(N),
\\
\hat T_F^{(A)}(N)&=&\hat 1_F(N),
\\
\quad\hat 1_F(N)&=&\displaystyle\sum_{\alpha_1<\beta_1<\cdots\alpha_N\beta_N}\vert\alpha_1\beta_1\cdots\alpha_N\beta_N\rangle\langle\alpha_1\beta_1\cdots\alpha_N\beta_N\vert,
\end{array}
\end{equation}
\begin{equation}
\begin{array}{lll}
\hat T_B^{(A)}&=&\displaystyle\sum_{N=0}^{N_{max}}\hat T_B(N),
\\
\hat T_B^{(A)}(N)&=&\displaystyle\sum_{\alpha_1<\beta_1<\cdots\alpha_N\beta_N}\vert\alpha_1\beta_1\cdots\alpha_N\beta_N)_M{}_M(\alpha_1\beta_1\cdots\alpha_N\beta_N\vert.
\end{array}
\end{equation}
$\hat Z^{(A)}$ is written as
\begin{equation}
\hat Z^{(A)}=(2\hat N_B^{(A)}-1)!!\hat T_B^{(A)}.
\end{equation}
The mapping operator is given as
\begin{equation}
\begin{array}{lll}
U_\xi^{(A)}&=&\displaystyle\sum_{N=0}^{N_{max}}U_\xi(N),
\\
U_\xi^{(A)}(N)&=&\displaystyle\{(2N-1)!!\}^\xi\sum_{\alpha_1<\beta_1<\cdots<\alpha_N<\beta_N}\vert\alpha_1\beta_1\cdots\alpha_N\beta_N)_M\ \langle\alpha_1\beta_1\cdots\alpha_N\beta_N\vert.
\end{array}
\end{equation}
If we set $\xi=0$ and $N_{max}\rightarrow\infty$, this mapping becomes the MYT mapping \cite{MYT64},
from which we obtain the boson expansions of Holstein and Primakoff, and if we take $\xi=\pm1$, they become mapping operators for the Dyson boson expansions \cite{JDF71}.
Taking $N_{max}\rightarrow\infty$, $U_\xi^{(A)}$ maps the whole fermion space that consists of even numbers of quasi-particles to the boson subspace.

\subsection{The case where the maximum phonon excitation number is 1}
In the case where the maximum phonon excitation number is 1, 
$\vert 0\rangle$ and $\vert\mu\rangle$ become orthonormal bases of the fermion space of even-quasi-particle excitations up to the two-quasi-particle excitations, 
which correspond to $\vert 0)$ and $\vert \mu)$, respectively.
The mapping operator becomes as
\begin{equation}
\label{eq:ubarmax1}
U_\xi=\widetilde U=\vert 0)\langle 0\vert+\sum_\mu\vert\mu)\langle \mu\vert,
\end{equation}
and $\hat Z=\breve 1_B$.
As a result, the mapping operator has no dependence on $\xi$, and the mapping becomes of the Hermitian type.
The projection operator onto the fermion subspace to be mapped is given by
\begin{subequations}
\label{eq:tftbmax1}
\begin{equation}
\label{eq:tfmax1}
\hat T_F=\vert 0\rangle\langle 0\vert +\sum_\mu\vert \mu\rangle\langle \mu\vert.
\end{equation}
The projection operator onto the physical subspace, which has a one-to-one correspondence to the fermion subspace, becomes as
\begin{equation}
\label{eq:tbmax1}
\hat T_B=\breve 1_B.
\end{equation}
\end{subequations}
This indicates that the ideal boson states are the physical state vectors, with one-to-one correspondences to the fermion state vectors.

The following relations,
\begin{subequations}
\label{eq:formulamax1}
\begin{equation}
\label{eq:formulaxtmax1}
X_\mu U_{-\xi}^\dagger =\vert 0\rangle(\mu\vert =U_{-\xi}^\dagger b_t\breve 1_B,
\end{equation}
\begin{equation}
\label{eq:formulaxtdaggermax1}
U_\xi X_\mu^\dagger =\vert \mu)\langle 0\vert =\breve 1_Bb_\mu^\dagger U_\xi,
\end{equation}
\begin{equation}
\label{eq:formulabqmax1}
B_qU_{-\xi}^\dagger =\sum_{\mu'}\sum_\mu\Gamma^{\mu'\mu}_q\vert\mu\rangle(\mu'\vert,
\end{equation}
\end{subequations}
hold, and we obtain
\begin{subequations}
\label{eq:bexpmax1}
\begin{equation}
\label{eq:bexpxtmax1}
(X_\mu)_\xi=\vert 0)(\mu\vert =\breve 1_B(X_\mu)_B\breve 1_B=(X_\mu)_B\breve 1_B,\quad (X_\mu)_B=b_\mu,
\end{equation}
\begin{equation}
\label{eq:bexpxtdaggermax1}
(X_\mu^\dagger)_\xi=\vert\mu)(0\vert =\breve 1_B(X_\mu^\dagger)_B\breve 1_B=\breve 1_B(X_\mu^\dagger)_B,\quad (X_\mu^\dagger)_B=b_\mu^\dagger,
\end{equation}
\begin{equation}
\label{eq:bexpbqmax1}
\begin{array}{c}
(B_q)_\xi=\displaystyle\sum_{\mu'}\sum_\mu\Gamma^{\mu'\mu}_q\vert\mu)(\mu'\vert=\breve 1_B(B_q)_B\breve 1_B=\breve 1_B(B_q)_B=(B_q)_B\breve 1_B,
\\
 (B_q)_B=\displaystyle\sum_{\mu \mu'}\Gamma^{\mu' \mu}_qb_\mu^\dagger b_{\mu'}.
\end{array}
\end{equation}
\end{subequations}

The product of the operators becomes as follows:
\begin{subequations}
\label{eq:bexpproductmax1}
\begin{equation}
(O_FX_\mu)_\xi=(O_F)_\xi(X_\mu)_\xi=\breve 1_B(O_F)_B\breve 1_B(X_\mu)_B\breve 1_B=\breve 1_B(O_F)_B(X_\mu)_B\breve 1_B,
\end{equation}
\begin{equation}
(X_\mu^\dagger O_F)_\xi=(X_\mu^\dagger)_\xi(O_F)_\xi=\breve 1_B (X_\mu^\dagger)_B\breve 1_B(O_F)_B\breve 1_B=\breve 1_B(X_\mu^\dagger)_B(O_F)_B\breve 1_B,
\end{equation}
\end{subequations}
therefore we can obtain the mapping of the product of $X_\mu^\dagger$,$X_\mu$, and $B_q$ by arranging them in normal order.

The commutation relations of $(X_\mu^\dagger )_B$, $(X_\mu)_B$, and $(B_q)_B$ become as follows:
\begin{subequations}
\label{eq:combapp}

\begin{equation}
\label{eq:combapp1}
[ (X_\mu)_B, (X_{\mu'}^\dagger)_B ]=\delta_{\mu,\mu'}
\end{equation}

\begin{equation}
\label{eq:combapp2}
[ (B_q)_B,( X_\mu^\dagger)_B]=\sum_{\mu'}\Gamma^{\mu\mu'}_q(X_{\mu'}^\dagger)_B.
\end{equation}

\begin{equation}
\label{eq:combapp3}
[ (X_\mu)_B, (B_q )_B]=\sum_{\mu'}\Gamma^{\mu'\mu}_q(X_{\mu'})_B,
\end{equation}
\end{subequations}
which are equal to the results of the boson approximation.

From the above, when the maximum number of phonons is $1$, by arranging the phonon creation and annihilation operators and the scattering operators in normal order and replacing them with $(X_\mu^\dagger)_B$, $(X_\mu)_B$, and  $(B_q)_B$, respectively, then the fermion subspace is completely mapped onto the boson subspace projected by $\breve 1_B$.

In this way, NOM establishes the boson approximation as the boson mapping whose maximum phonon excitation number is 1.

\section{Boson expansions}
\label{bexp}

\subsection{Formulae for the boson expansions}
We give here the formulae used to obtain the boson expansions of the mapped fermion operators.
We utilize
\begin{equation}
\begin{array}{lll}
\label{eq:overlineunexp}
\displaystyle\widetilde{U}(N)&=&\displaystyle\sum_{t_1\leq t_2\leq\cdots\leq t_N}\vert t_1 t_2\cdots t_N)\langle t_1t_2\cdots t_N\vert
\\
&=&\displaystyle\sum_{t_1t_2\cdots t_N}\frac{{\mathcal N}_B(t_1t_2\cdots t_N)}{N!}\vert t_1 t_2\cdots t_N)\langle t_1t_2\cdots t_N\vert
\\
&=&\displaystyle\sum_{t_1t_2\cdots t_N}\frac{1}{N!}\vert t_1 t_2\cdots t_N))\langle \langle t_1t_2\cdots t_N\vert
\end{array}
\end{equation}
and obtain the following series of formulae:
\begin{subequations}
\label{eq:ubarop}
\begin{equation}
\widetilde{U}(N)X_{t'}=(X_{t'})_D\widetilde{U}(N+1)\quad (N\geq 0),
\end{equation}
\begin{equation}
\begin{array}{lll}
\widetilde{U}(1)X_t^\dagger&=&(X_t^\dagger)_D\widetilde{U}(0),
\\
\widetilde{U}(N+1)X_t^\dagger&=&(X_t^\dagger)_D\widetilde{U}(N)
\\
&&-\displaystyle\frac12\sum_{t_1t_2}\sum_{\bar t_1'}Y(tt_1t_2\bar t'_1)b_{t_1}^\dagger b_{t_2}^\dagger\widetilde{U}(N-1)X_{\bar t'_1}\quad (N\geq 1),
\end{array}
\end{equation}

\begin{equation}
\begin{array}{lll}
\widetilde{U}(0)B_q&=&0,
\\
\widetilde{U}(N)B_q&=&(B_q)_D\widetilde{U}(N)+\displaystyle\sum_t\sum_{\bar t'}\Gamma_q^{\bar t't}b_t^\dagger\widetilde{U}(N-1)X_{\bar t'}\quad (N\geq 1),
\end{array}
\end{equation}

\begin{equation}
\begin{array}{lll}
\widetilde{U}(1)X_{\bar t}^\dagger&=&0,
\\
\widetilde{U}(N+1)X_{\bar t}^\dagger&=&\displaystyle -\frac12\sum_{t_1t_2}\sum_{t'_1}Y(\bar t t_1t_2t'_1)b_{t_1}^\dagger b_{t_2}^\dagger b_{t'_1}\widetilde{U}(N)
\\
&&\displaystyle -\frac 12 \sum_{t_1t_2}\sum_{\bar t'_1}Y(\bar t t_1t_2\bar t'_1)b_{t_1}^\dagger b_{t_2}^\dagger \widetilde{U}(N-1)X_{\bar t'_1}\quad (N\geq 1),
\end{array}
\end{equation}
\end{subequations}

where
\begin{subequations}
\label{eq:dbexp}
\begin{equation}
(X_{t'})_D=b_{t'},
\end{equation}
\begin{equation}
\label{eq:xdaggerd}
(X_t^\dagger)_D=b_t^\dagger-\frac 12\sum_{t_1t_2}\sum_{t'_1}Y(tt_1t_2t'_1)b_{t_1}^\dagger b_{t_2}^\dagger b_{t'_1},
\end{equation}
\begin{equation}
\label{eq:bqd}
(B_q)_D=\sum_t\sum_{t'}\Gamma_q^{t't}b_t^\dagger b_{t'}.
\end{equation}
\end{subequations}
Eqs. (\ref{eq:dbexp}) are the same as the boson expansions derived by DBET.  $(B_{\bar q})_D^\dagger =(B_q)_D$ holds.

From these formulae, we obtain
\begin{subequations}
\label{eq:uoun}

\begin{equation}
\widetilde{X_{t'}}(N) =(X_{t'})_D\hat Z(N+1)\qquad (N\geq 0),
\end{equation}

\begin{equation}
\label{eq:barxtdagn}
\begin{array}{lll}
\widetilde{X_t^\dagger}(0)&=&(X_t^\dagger )_D\hat Z(0)
\\
\widetilde{X_t^\dagger}(N)&=&(X_t^\dagger )_D\hat Z(N)
 -\displaystyle\frac 12\sum_{t_1t_2}\sum_{\bar t'_1}Y(tt_1t_2\bar t'_1)b_{t_1}^\dagger b_{t_2}^\dagger \widetilde{X_{\bar t'_1}}(N-1)\qquad (N\ge 1),
\end{array}
\end{equation}

\begin{equation}
\label{eq:barbqn}
\begin{array}{lll}
\widetilde{B_q}(0)&=&0,
\\
\widetilde{B_q}(N) &=&(B_q)_D\hat Z(N)+\displaystyle\sum_t\sum_{\bar t'}\Gamma_q^{\bar t't}b_t^\dagger\widetilde{X}_{\bar t'}(N-1)\qquad (N\ge 1),
\end{array}
\end{equation}

\begin{equation}
\label{eq:barxbtdagn}
\begin{array}{lll}
\widetilde{X_{\bar t}^\dagger}(0)&=&0,
\\
\widetilde{X_{\bar t}^\dagger}(N)
&=&-\displaystyle\frac 12\sum_{t_1t_2}\sum_{t'_1}Y(\bar tt_1t_2t'_1)b_{t_1}^\dagger b_{t_2}^\dagger b_{t'_1}\hat Z(N)
\\
&-&\displaystyle\frac 12\sum_{t_1t_2}\sum_{\bar t'_1}Y(\bar tt_1t_2\bar t'_1)b_{t_1}^\dagger b_{t_2}^\dagger \widetilde{X_{\bar t'_1}}(N-1)\qquad (N\ge 1),
\end{array}
\end{equation}
\end{subequations}
where we use the following diffinitions: $\widetilde{X_\mu}(N)=\widetilde U(N)X_\mu\widetilde U(N+1)^\dagger$, $\widetilde{B_q}(N)=\widetilde U(N)B_q\widetilde U(N)^\dagger$. $\widetilde{X_\mu^\dagger}(N)=\left (\widetilde{X_\mu}(N)\right )^\dagger$.
$\widetilde{B_q^\dagger}(N)=\left ({\widetilde B_q}(N)\right )^\dagger$ holds.

We can obtain the boson expansion of $\hat Z(N)$ by using 
\begin{equation}
\label{eq:zn}
\hat Z(N)=\frac 1N\sum_t(X_t^\dagger)_D\hat Z(N-1)b_t-\frac 1{2N}\sum_{t_1t_2}\sum_{t'_1}\sum_{\bar t'}Y(t'_1t_1t_2\bar t')b_{t_1}^\dagger b_{t_2}^\dagger\widetilde{X_{\bar t'}}(N-2) b_{t'_1}\quad (N\geq 2),
\end{equation}
which is derived by applying
\begin{equation}
\widetilde{U}^\dagger (N) =\frac 1N\sum_tX_t^\dagger\widetilde{U}^\dagger (N-1)b_t\qquad (N\ge 1)
\end{equation}
obtained from Eq. (\ref{eq:overlineunexp}) to Eq. (\ref{eq:tldeurel1}), expressing $\hat Z(N)$ as
\begin{equation}
\label{eq:znf}
\hat Z(N)=\frac 1N\sum_t\widetilde{X_t^\dagger}(N-1)b_t\quad (N\geq 1),
\end{equation}
and substituting Eqs. (\ref{eq:barxtdagn}) into this.
$\hat Z(N)$ up to $N=2$ are as follows,
\begin{subequations}
\label{eq:z012}
\begin{equation}
\label{eq:z0}
\hat Z(0)=\hat 1_B(0), 
\end{equation}
\begin{equation}
\label{eq:z1}
\hat Z(1)=\hat 1_B(1), 
\end{equation}
\begin{equation}
\label{eq:z2}
\hat Z(2)=\hat 1_B(2)\left (\hat 1_B-\frac 14\sum_{t_1t_2}\sum_{t'_1t'_2}Y(t'_2t_1t_2t'_1)b_{t_1}^\dagger b_{t_2}^\dagger b_{t'_1}b_{t'_2}\right )\hat 1_B(2).
\end{equation}
\end{subequations}
We use the following equation for the case $N=2$,
\begin{equation}
\label{eq:b1bn}
b_t\hat1_B(N)=\hat 1_B(N-1)b_t.
\end{equation}

Once $\widetilde{X_{\bar t'}}(N)$, $\widetilde{X_{\bar t}}^\dagger(N)$, and $\hat Z(N)$ are obtained from Eq. (\ref{eq:barxbtdagn}), the Hermitian conjugate of Eq. (\ref{eq:barxbtdagn}), and Eq. (\ref{eq:zn}),  $\widetilde{X_{t'}}(N)$, $\widetilde{X_t}^\dagger (N)$, and $\widetilde{B_q}(N)$ are given by substituting these into Eqs. (\ref{eq:uoun}).

\subsection{On the use of ideal boson state vectors}
The effect of the Pauli exclusion principle is reflected generally in the boson operators and the boson state vectors by the mapping. While, if we restrict the types of phonon excitation modes and the number of phonon excitations so that zero eigenvalues do not appear in the norm matrices of the multiphonon state vectors, then
\begin{equation}
\hat T_B=\breve 1_B,
\end{equation}
holds.
In this case, the ideal boson state vectors $\vert N; t)$, which do not bear the effect of the Pauli exclusion principle, become the physical state vectors.
As a result, all effects of the Pauli exclusion principle are fully reflected in the mapped operators.

In order that the boson expansion method is practical, the phonon excitation modes and the maximum number of excitations should be chosen so that the ideal boson state vectors become the physical state vectors \cite{KT83}.

\subsection{Boson expansions as a small parameter expansion}
\label{smallpexp}
In this subsection, we obtain the norm operator and the other mapped operators in the boson expansion being a small parameter expansion,
where $\Gamma_q^{\mu\mu'}$ are regarded as of the order of magnitude $O(\Gamma)$.

\subsubsection{On the conditions for being a small parameter expansion and the evaluation of the order of magnitude of each term of expansions}
For realizing a small parameter expansion where the boson approximation becomes the zeroth order approximation, $\hat Z\approx\breve 1_B$ must hold as the zeroth order approximation. 
For that purpose, it is necessary to limit the type of mode and the number of phonon excitations in the mapping operator so that zero eigenvalues do not appear in the norm matrices of the multiphonon state vectors. 
This is the same condition for the ideal boson state vectors to become physical.
This condition is necessary but not sufficient, however.

Denoting the matrix  each element of which is $\langle t'_1t'_2\vert t_1t_2\rangle$ as $\mathbf{Z}(N)$, $\mathbf{Z}^{(A)}(N)$ is expressed  as
\begin{equation}
\label{eq:zwwtz'}
\mathbf{Z}^{(A)}(N)=\left (
\begin{array}{@{\,}cc@{\,}}
\mathbf{Z}(N)&\mathbf{W}(N)
\\
\mathbf{W}(N)^T&\mathbf{Z'}(N).
\end{array}
\right ),
\end{equation}
 As shown in the appendix, if $\mathbf{W}(2)=\mathbf{0}(2)$, i.e. $Y(t'_1\bar t\mu t'_2)=0$, then $\mathbf{W}(N)=\mathbf{0}(N)$ for $N\geq 3$.
Hence, in this case, we obtain
\begin{equation}
\mathbf{Z}^{(A)}(N)=\left (
\begin{array}{@{\,}cc@{\,}}
\mathbf{Z}(N)&\mathbf{0}(N)
\\
\mathbf{0} (N)^T&\mathbf{Z'}(N)
\end{array}
\right ).
\end{equation}
Substituting this into Eq.(\ref{eq:znarelmt}),
\begin{subequations}
\label{eq:znarelmtzz'}
\begin{equation}
\label{eq:znarelmtz}
\mathbf{Z}(N)^2=(2N-1)!!\mathbf{Z}(N),
\end{equation}
\begin{equation}
\label{eq:znarelmtz'}
\mathbf{Z'}(N)^2=(2N-1)!!\mathbf{Z'}(N),
\end{equation}
\end{subequations}
are obtained.
From Eq. (\ref{eq:znarelmtz}), we obtain
\begin{equation}
\hat Z(N)^2=(2N-1)!!\hat Z(N),
\end{equation}
from which we find
\begin{subequations}
\label{eq:znnons}
\begin{equation}
\label{eq:znnons1}
\hat Z(N)=(2N-1)!!\hat T_B(N)=(2\hat N_B-1)!!\hat T_B(N),\quad \hat N_B=\sum_tb_t^\dagger b_t,
\end{equation}
\begin{equation}
\label{eq:znnons2}
\hat Z=(2\hat N_B-1)!!\hat T_B.
\end{equation}
\end{subequations}
$\{t\}$ and $N_{max}$ are set so that no zero eigenvalue appears in $\mathbf{Z}(N)$.  It is $\mathbf{Z'}(N)$ that has zero eigenvalues.
Therefore the eigenvalues of $\mathbf{Z}(N)$ are only $(2N-1)!!$.
$\hat T_B(N)=\hat 1_B(N)$, and then $\hat T_B=\breve 1_B$ hold.
Even so, $\hat Z\approx\breve 1_B$ does not hold as the zeroth order approximation, and the boson expansions can not be obtained as the small parameter expansion.
$\mathbf{W}(2)$ must not be a zero matrix if the small parameter expansion holds.

We investigate the case of N=2 to establish the small parameter expansion and an order evaluation of the terms in them.
Substituting Eq. (\ref{eq:zwwtz'}) into Eq. (\ref{eq:znarelmt}) and taking $N=2$, we obtain
\begin{subequations}
\label{relzwz'}
\begin{equation}
\label{relzwz'1}
\mathbf{Z}(2)^2+\mathbf{W}(2)\mathbf{W}(2)^T=3\mathbf{Z}(2),
\end{equation}
\begin{equation}
\label{relzwz'2}
\mathbf{Z}(2)\mathbf{W}(2)+\mathbf{W}(2)\mathbf{Z'}(2)=3\mathbf{W}(2),
\end{equation}
\begin{equation}
\label{relzwz'3}
\mathbf{W}(2)^T\mathbf{W}(2)+\mathbf{Z'}(2)^2=3\mathbf{Z'}(2),
\end{equation}
\end{subequations}
from which we derive
\begin{subequations}
\label{eq:yw}
\begin{equation}
\label{eq:yw1}
\sum_{\mu\mu'}Y(t'_1\mu\mu't'_2)Y(t_1\mu\mu't_2)=4((t'_1t'_2\vert t_1t_2))-2Y(t'_1t_1t_2t'_2),
\end{equation}
\begin{equation}
\label{eq:yw2}
\sum_{\mu\mu'}Y(t'_1\mu\mu't'_2)Y(\bar t_1\mu\mu'\mu_1)+2Y(t'_1\bar t_1\mu_1t'_2)=0,
\end{equation}
\begin{equation}
\label{eq:yw3}
\sum_{\mu\mu'}Y(\bar t'_1\mu\mu'\mu'_1)Y(\bar t_1\mu\mu'\mu_1)=4((\bar t'_1\mu'_1\vert\bar t_1\mu_1))-2Y(\bar t'_1\bar t_1\mu_1\mu'_1),
\end{equation}
\end{subequations}
Since $\mathbf{Z}^{(A)}(2)$ has zero eigenvalues \cite{KT83}, these relations include some parts where the small parameter expansion breaks down.
$Y(\mu_1\mu_2\mu_3\mu_4)\sim O(\Gamma^2)$ should hold. Therefore if $\mu$-sums do not affect the evaluation of the order of magnitude,  these equations have discrepancies in the order of magnitude of each term. The naive evaluation does not hold, and we must correctly evaluate the case where we take $\mu$-sum.

We choose $\{t\} $ so that the small parameter expansion holds in any situation. $\displaystyle\sum_{tt'}Y(t_1tt't_2)Y(t'_1tt't'_2)$ should, then, be estimated as $O(\Gamma^4)$.
To find out more about $\bar t$-sum, we take up $\displaystyle\sum_\mu Y(t_1t_2t_3\mu)\Gamma_q^{\mu t_4}$.  Because we choose $\{t\}$ so that $\sum_tY(t_1t_2t_3t)\Gamma_q^{tt_4}\sim O(\Gamma^3)$ hold, then we obtain
\begin{equation}
\sum_\mu Y(t_1t_2t_3\mu)\Gamma_q^{\mu t_4}=\sum_{\bar t}Y(t_1t_2t_3\bar t)\Gamma_q^{\bar tt_4}+O(\Gamma^3).
\end{equation}
While
\begin{equation}
\label{eq:musumyg}
\sum_\mu Y(t_1t_2t_3\mu)\Gamma_q^{\mu t_4}=\sum_{q'q''}\sum_{\alpha\beta\gamma}\varphi_q(\alpha\beta)\varphi_{q'}(\gamma\alpha)\varphi_{q''}(\gamma\alpha)(\Gamma_{q'}^{t_1t_2}\Gamma_{q''}^{t_3t_4}+\Gamma_{q'}^{t_1t_3}\Gamma_{q''}^{t_2t_4}),
\end{equation}
holds, which indicates that the order of the right-hand side is $O(\Gamma^2)$. Therefore the estimation of  $\bar t$-sum should become as
\begin{equation}
\label{eq:musummygordr}
\sum_{\bar t} Y(t_1t_2t_3\bar t)\Gamma_q^{\bar tt_4}\sim O(\Gamma^2).
\end{equation}
This indicates that if we take a single $\bar t$-sum, we should estimate its magnitude by one order lower.
Based on this evaluation, we evaluate
\begin{equation}
\sum_{t\bar t}Y(t_1t\bar tt_2)Y(t'_1t\bar tt'_2)\sim O(\Gamma^3),
\end{equation}
By applying these order evaluations to Eqs. (\ref{eq:yw}), we obtain
\begin{subequations}
\label{eq:spestres}
\begin{equation}
\label{eq:spestres1}
\sum_{\bar t\bar t'}Y(\mu'_1\bar t\bar t'\mu'_2)Y(\mu_1\bar t\bar t'\mu_2)=4((\mu'_1\mu'_2\vert\mu_1\mu_2))-2Y(\mu'_1\mu_1\mu_2\mu'_2)+O(\Gamma^3),
\end{equation}
\begin{equation}
\label{eq:spestres2}
\sum_{\bar t\bar t'}Y(t'_1\bar t\bar t't'_2)Y(\bar t_1\bar t\bar t'\mu_1)+2Y(t'_1\bar t_1\mu_1t'_2)=O(\Gamma^3),
\end{equation}
\begin{equation}
\label{eq:spestres3}
\sum_{\bar t\bar t'}Y(\bar t'_1\bar t\bar t'\mu'_1)Y(\bar t_1\bar t\bar t'\mu_1)=4((\bar t'_1\mu'_1\vert\bar t_1\mu_1))-2Y(\bar t'_1\bar t_1\mu_1\mu'_1)+O(\Gamma^3).
\end{equation}
\end{subequations}
We can identify that the parts where the double $\bar t$-sums are performed across two coefficients are responsible for the failure of the small parameter expansion. 
Eqs. (\ref{eq:spestres}) become conditions for the small parameter expansion to hold.

\subsubsection{Boson expansions of mapped operators as the small parameter expansion}
Here we perform the boson expansions of the mapped operators as the small parameter expansion.

Eq. (\ref{eq:ximapzbar2}) indicates that we can derive the boson expansions of $(O_F)_\xi$ from those of the norm operator $\hat Z$ and $\widetilde{O_F}$. We give the terms of the boson expansions up to the order of magnitude $O(\Gamma^4)$.

From Eq. (\ref{eq:barxbtdagn}), its Hermitian conjugate, and Eq. (\ref{eq:zn}), we find the recurrence formulae for obtaining the boson expansions of $\hat Z(N)$,
$\widetilde{X_{\bar t'}}(N)$, and $\widetilde{X_{\bar t}^\dagger}(N)$ up to the desired order of magnitude.
These recurrence formulae generate no parts where double $\bar t$-sums are performed across two coefficients in the expansions, which makes it possible to avoid convergence difficulty caused by them.
The recurrence formulae of $\hat Z(N)$ are as follows:
\begin{subequations}
\label{eq:znrec}
\begin{equation}
\label{eq:znrec01234}
\hat Z(N)=\sum_{k=1}^4\hat Z^{(k)}(N)+O(\Gamma^5);\quad \hat Z^{(k)}(N)\sim O(\Gamma^k),
\end{equation}
\begin{equation}
\label{eq:znrec0}
\hat Z^{(0)}(N)=\frac 1N\sum_tb_t^\dagger\hat Z^{(0)}(N-1)b_t,
\end{equation}
\begin{equation}
\label{eq:znrec1}
\hat Z^{(1)}(N)=0,
\end{equation}

\begin{equation}
\label{eq:znrec2}
\begin{array}{lll}
\hat Z^{(2)}(N)&=&\displaystyle\frac 1N\sum_tb_t^\dagger\hat Z^{(2)}(N-1)b_t
\\
&&-\displaystyle\frac 1{2N}\sum_{t_1t_2}\sum_{t'_1t'_2}Y(t'_2t_1t_2t'_1)b_{t_1}^\dagger b_{t_2}^\dagger b_{t'_1}\hat Z^{(0)}(N-1)b_{t'_2},
\end{array}
\end{equation}
\begin{equation}
\label{eq:znrec3}
\begin{array}{lll}
\hat Z^{(3)}(N)&=&\displaystyle\frac 1N\sum_tb_t^\dagger\hat Z^{(3)}(N-1)b_t
\\
&&+\displaystyle\frac 1{4N}\sum_{t_1t_2t_3}\sum_{t'_1t'_2t'_3}\sum_{\bar t}Y(t'_3t_1t_2\bar t)Y(\bar tt'_1t'_2t_3)b_{t_1}^\dagger b_{t_2}^\dagger\hat Z^{(0)}(N-1)b_{t_3}^\dagger b_{t'_1}b_{t'_2}b_{t'_3},
\end{array}
\end{equation}

\begin{equation}
\label{eq:znrec4}
\begin{array}{c}
\hat Z^{(4)}(N)=\displaystyle\frac 1N\sum_tb_t^\dagger\hat Z^{(4)}(N-1)b_t
-\displaystyle\frac 1{2N}\sum_{t_1t_2}\sum_{t'_1t'_2}Y(t'_2t_1t_2t'_1)b_{t_1}^\dagger b_{t_2}^\dagger b_{t'_1}\hat Z^{(2)}(N-1)b_{t'_2}
\\
-\displaystyle\frac 1{8N}\sum_{t_1t_2t_3t_4}\sum_{t'_1t'_2t'_3t'_4}\sum_{\bar t\bar t'}Y(t'_4t_1t_2\bar t)Y(\bar tt'_2t'_3\bar t')Y(\bar t't_3t_4t'_1)b_{t_1}^\dagger b_{t_2}^\dagger b_{t_3}^\dagger b_{t_4}^\dagger b_{t'_1}\hat Z^{(0)}(N-1)b_{t'_2}b_{t'_3}b_{t'_4}.
\end{array}
\end{equation}
\end{subequations}

The solution of Eq,(\ref{eq:znrec0}) is easily obtained as
\begin{equation}
\label{eq:znrec0sol}
\hat Z^{(0)}(N)=\frac 1{N!}\sum_{t_1t_2\cdots t_N}b_{t_1}^\dagger b_{t_2}^\dagger\cdots b_{t_N}^\dagger\hat Z(0)b_{t_1}b_{t_2}\cdots b_{t_N}=\hat 1_B(N).
\end{equation}
Substituting it into Eq. (\ref{eq:znrec2}) and using Eq. (\ref{eq:b1bn}),
we obtain
\begin{subequations}
\begin{equation}
\label{eq:znrec2'}
\begin{array}{lll}
\hat Z^{(2)}(N)&=&\displaystyle\frac 1N\sum_tb_t^\dagger\hat Z^{(2)}(N-1)b_t
\\
&&-\displaystyle\frac 1{2N}\sum_{t_1t_2}\sum_{t'_1t'_2}Y(t'_2t_1t_2t'_1)b_{t_1}^\dagger b_{t_2}^\dagger\hat 1_B(N-2) b_{t'_1}b_{t'_2}.
\end{array}
\end{equation}
For finding the solution, assuming it as
\begin{equation}
\hat Z^{(2)}(N)=y^{(2)}(N)\sum_{t_1t_2}\sum_{t'_1t'_2}Y(t'_2t_1t_2t'_1)b_{t_1}^\dagger b_{t_2}^\dagger\hat 1_B(N-2) b_{t'_1}b_{t'_2},
\end{equation}
substituting this into the recurrence formula, and using 
\begin{equation}
\label{eq:bd1bb}
\sum_tb_t^\dagger\hat 1_B(N)b_t=(N+1)\hat 1_B(N+1),
\end{equation}
we find
\begin{equation}
y^{(2)}(N)=\frac{N-2}Ny^{(2)}(N-1)-\frac1{2N}.
\end{equation}
$y^{(2)}(N)=-1/4$ is the solution, and we obtain
\begin{equation}
\hat Z^{(2)}(N)=-\frac 14\sum_{t_1t_2}\sum_{t'_1t'_2}Y(t'_2t_1t_2t'_1)b_{t_1}^\dagger b_{t_2}^\dagger\hat 1_B(N-2) b_{t'_1}b_{t'_2}.
\end{equation}
\end{subequations}
Following the same procedure in order, we can obtain the solution of the recurrence formulae for each order of magnitude.
Organizing the solutions obtained in this way using Eq. (\ref{eq:b1bn}), we finally obtain
\begin{subequations}
\label{eq:znexp}
\begin{equation}
\label{eq:znor4}
\hat Z(N)=\mathcal{\hat Z}\hat 1_B(N)=\hat 1_B(N)\mathcal{\hat Z}=\hat 1_B(N)\mathcal{\hat Z}\hat 1_B(N),
\end{equation}
\begin{equation}
\mathcal{\hat Z}=\mathcal{\hat Z}^{(0)}+\mathcal{\hat Z}^{(2)}+\mathcal{\hat Z}^{(3)}+\mathcal{\hat Z}^{(4)}+O(\Gamma^5),
\end{equation}
\begin{equation}
\label{eq:zn0}
\mathcal{\hat Z}^{(0)}=\hat 1_B,
\end{equation}
\begin{equation}
\label{eq:zn2}
\mathcal{\hat Z}^{(2)}=-\frac 14\sum_{t_1t_2}\sum_{t'_1t'_2}Y(t'_2t_1t_2t'_1)b_{t_1}^\dagger b_{t_2}^\dagger b_{t'_1}b_{t'_2},
\end{equation}

\begin{equation}
\label{eq:zn3}
\mathcal{\hat Z}^{(3)}=\frac 1{12}\sum_{t_1t_2t_3}\sum_{t'_1t'_2t'_3}\sum_{\bar t}Y(t'_3t_1t_2\bar t)Y(\bar tt'_1t'_2t_3)b_{t_1}^\dagger b_{t_2}^\dagger b_{t_3}^\dagger b_{t'_1}b_{t'_2}b_{t'_3},
\end{equation}

\begin{equation}
\label{eq:zn4}
\begin{array}{lll}
\mathcal{\hat Z}^{(4)}&=&\mathcal{\hat Z}_{in}^{(4)}+\mathcal{\hat Z}_{out}^{(4)}
\\
&\mathcal{\hat Z}_{in}^{(4)}&=\displaystyle\frac 1{12}\sum_{t_1t_2t_3}\sum_{t'_1t'_2t'_3}\sum_{t}Y(t'_1t_1t_2 t)Y(tt'_2t'_3t_3)b_{t_1}^\dagger b_{t_2}^\dagger b_{t_3}^\dagger b_{t'_1}b_{t'_2}b_{t'_3}
\\
&&+\displaystyle\frac 1{32}\sum_{t_1t_2t_3t_4}\sum_{t'_1t'_2t'_3t'_4}Y(t'_2t_1t_2t'_1)Y(t'_4t_3t_4t'_3)b_{t_1}^\dagger b_{t_2}^\dagger b_{t_3}^\dagger b_{t_4}^\dagger b_{t'_1}b_{t'_2}b_{t'_3}b_{t'_4},
\\
&\mathcal{\hat Z}_{out}^{(4)}&=-\displaystyle\frac 1{32}\sum_{t_1t_2t_3t_4}\sum_{t'_1t'_2t'_3t'_4}\sum_{\bar t\bar t'}Y(t'_4t_1t_2\bar t)Y(\bar tt'_2t'_3\bar t')Y(\bar t't_3t_4t'_1)b_{t_1}^\dagger b_{t_2}^\dagger b_{t_3}^\dagger b_{t_4}^\dagger b_{t'_1}b_{t'_2}b_{t'_3}b_{t'_4}.
\end{array}
\end{equation}
\end{subequations}
From these results, we can easily find the norm operator as
\begin{equation}
\label{eq:normopres}
\begin{array}{lll}
\hat Z&=&\mathcal{\hat Z}\breve 1_B=\breve 1_B\mathcal{\hat Z}=\breve 1_B\mathcal{\hat Z}\breve 1_B,
\\
&\mathcal{\hat Z}&=\hat 1_B+\mathcal{\hat Y},
\\
&&\mathcal{\hat Y}=\mathcal{\hat Y}_{in}+\mathcal{\hat Y}_{out},
\\
&&\quad\mathcal{\hat Y}_{in}=\mathcal{\hat Z}^{(2)}+\mathcal{\hat Z}_{in}^{(4)}+O(\Gamma^5),
\\
&&\quad\mathcal{\hat Y}_{out}=\mathcal{\hat Z}^{(3)}+\mathcal{\hat Z}_{out}^{(4)}+O(\Gamma^5).
\end{array}
\end{equation}
The $\xi$-th power of $\hat Z$ becomes
\begin{equation}
\label{eq:normopresp}
\begin{array}{l}
\hat Z^\xi=\mathcal{\hat Z}^\xi\breve 1_B=\breve 1_B\mathcal{\hat Z}^\xi,
\\
\quad\mathcal{\hat Z}^\xi=\hat 1_B+\xi\mathcal{\hat Y}+\displaystyle\frac 12\xi(\xi-1)\mathcal{\hat Y}^2+O(\Gamma^6).
\end{array}
\end{equation}

Once $\hat Z(N)$ is known, we can obtain $\widetilde{X_{\bar t}}(N)$ from the following recurrence formula derived from Eqs. (\ref{eq:barxbtdagn}),
\begin{equation}
\begin{array}{lll}
\widetilde{X_{\bar t'}}(N)&=&-\displaystyle\frac 12\sum_{t_1}\sum_{t'_1t'_2}Y(\bar t't'_1t'_2t_1)\hat Z(N)b_{t_1}^\dagger b_{t'_1}b_{t'_2}
\\
&&+\displaystyle\frac 14\sum_{t_1t_2}\sum_{t'_1t'_2t'_3}\sum_{\bar t}Y(\bar t't'_2t'_3\bar t)Y(\bar t t_1t_2t'_1)b_{t_1}^\dagger b_{t_2}^\dagger b_{t'_1}\hat Z(N-1)b_{t'_2}b_{t'_3}
\\
&&+\displaystyle\frac 14\sum_{t_1t_2}\sum_{t'_1t'_2}\sum_{\bar t}\sum_{\bar t'_1}Y(\bar t't'_1t'_2\bar t)Y(\bar tt_1t_2\bar t'_1)b_{t_1}^\dagger b_{t_2}^\dagger\widetilde{X_{\bar t'_1}}(N-2)b_{t'_1}b_{t'_2},
\end{array}
\end{equation}
and find the solutions as
\begin{subequations}
\begin{equation}
\label{eq:olxbtN}
\widetilde{X_{\bar t'}}(N)=(X_{\bar t'})_L\mathcal{\hat Z}\hat1_B(N+1)=\hat1_B(N)(X_{\bar t'})_L\mathcal{\hat Z},
\end{equation}
\begin{equation}
 (X_{\bar t'})_L
= (X_{\bar t'})_L^{(2)}+(X_{\bar t'})_L^{(3)}+(X_{\bar t'})_L^{(4)}+O(\Gamma^5),
\end{equation}
\begin{equation}
\begin{array}{lll}
(X_{\bar t'})_L^{(2)}&=&-\displaystyle\frac 12\sum_{t_1}\sum_{t'_1t'_2}Y(\bar t't'_1t'_2t_1)b_{t_1}^\dagger b_{t'_1}b_{t'_2},
\\
(X_{\bar t'})_L^{(3)}&=&\displaystyle\frac 14\sum_{t_1t_2}\sum_{t'_1t'_2t'_3}\sum_{\bar t}Y(\bar t't'_2t'_3\bar t)Y(\bar tt_1t_2t'_1)b_{t_1}^\dagger b_{t_2}^\dagger b_{t'_1}b_{t'_2}b_{t'_3},
\\
(X_{\bar t'})_L^{(4)}&=&[\mathcal{\hat Z}^{(2)}, (X_{\bar t'})_L^{(2)}]
\\
&&-\displaystyle\frac18\sum_{t_1t_2t_3}\sum_{t'_1t'_2t'_3t'_4}\sum_{\bar t\bar t''}Y(\bar t't'_1t'_2\bar t)Y(\bar tt_1t_2\bar t'')Y(\bar t''t'_3t'_4t_3)b_{t_1}^\dagger b_{t_2}^\dagger b_{t_3}^\dagger b_{t'_1}b_{t'_2}b_{t'_3}b_{t'_4},
\end{array}
\end{equation}
\begin{equation}
\begin{array}{l}
[\mathcal{\hat Z}^{(2)}, (X_{\bar t'})_L^{(2)}]=-\displaystyle\frac 14\sum_{t_1}\sum_{t'_1t'_2}\sum_{tt'}Y(\bar t'tt't_1)Y(t'_1tt't'_2)b_{t_1}^\dagger b_{t'_1}b_{t'_2}
\\
\qquad-\displaystyle\frac 14\sum_{t_1t_2}\sum_{t'_1t'_2t'_3}\sum_t\left\{2Y(t_1t'_1t'_2t)Y(t\bar t't_2t'_3)-Y(\bar t't'_1t'_2t)Y(tt_1t_2t'_3)\right\}b_{t_1}^\dagger b_{t_2}^\dagger b_{t'_1}b_{t'_2}b_{t'_3}.
\end{array}
\end{equation}
\end{subequations}
From these, we obtain
\begin{equation}
\widetilde{X_{\bar t'}}=\sum_{N=0}^{N_{max}-1}\widetilde{X_{\bar t'}}(N)=(X_{\bar t'})_L\mathcal{\hat Z}\breve1_B=\breve 1_B^{(-1)}(X_{\bar t'})_L\mathcal{\hat Z}=\breve 1_B(X_{\bar t'})_L\mathcal{\hat Z}\breve 1_B,
\end{equation}
where
\begin{equation}
\breve 1_B^{(\Delta N)}=\sum_{N=0}^{N_{max}+\Delta N}\hat 1_B(N).
\end{equation}
$\breve 1_B^{(0)}=\breve 1_B$ and $\breve 1_B\breve 1_B^{(-1)}=\breve 1_B^{(-1)}\breve 1_B=\breve 1_B^{(-1)}$ hold.
Organizing the Hermitian conjugate of Eq. (\ref{eq:olxbtN}), we find
\begin{subequations}
\begin{equation}
\begin{array}{lll}
\widetilde{X_{\bar t}^\dagger}(N)&=&((X_{\bar t})_L\mathcal{\hat Z}\hat1_B(N+1))^\dagger
=\hat1_B(N+1)\mathcal{\hat Z}(X_{\bar t})_L^\dagger
\\
&=&\mathcal{\hat Z}(X_{\bar t})_L^\dagger\hat1_B(N)
=\left\{\mathcal{\hat Z}(X_{\bar t})_L^\dagger\mathcal{\hat Z}^{-1}\right\}\mathcal{\hat Z}\hat 1_B(N)
\\
&=&(X_{\bar t}^\dagger)_L\mathcal{\hat Z}\hat 1_B(N),
\end{array}
\end{equation}
where
\begin{equation}
(X_{\bar t}^\dagger)_L=\mathcal{\hat Z}(X_{\bar t})_L^\dagger\mathcal{\hat Z}^{-1},
\end{equation}
\begin{equation}
 (X_{\bar t}^\dagger)_L
=(X_{\bar t}^\dagger)_L^{(2)}+(X_{\bar t}^\dagger)_L^{(3)}+(X_{\bar t}^\dagger)_L^{(4)}+O(\Gamma^5),
\end{equation}
\begin{equation}
\begin{array}{lll}
(X_{\bar t}^\dagger)_L^{(2)}&=&((X_{\bar t})_L^{(2)})^\dagger,\quad (X_{\bar t}^\dagger)_L^{(3)}=((X_{\bar t})_L^{(3)})^\dagger,
\\
(X_{\bar t}^\dagger)_L^{(4)}&=&((X_{\bar t})_L^{(4)})^\dagger+[\mathcal{\hat Z}^{(2)}, ((X_{\bar t})_L^{(2)})^\dagger]
\\
&=& -\displaystyle\frac 18\sum_{t_1t_2t_3t_4}\sum_{t'_1t'_2t'_3}\sum_{\bar t'\bar t''}Y(\bar tt_1t_2\bar t')Y(\bar t't'_1t'_2\bar t'')Y(\bar t''t_3t_4t'_3)b_{t_1}^\dagger b_{t_2}^\dagger b_{t_3}^\dagger b_{t_4}^\dagger b_{t'_1} b_{t'_2} b_{t'_3},
\end{array}
\end{equation}
\end{subequations}
and obtain
\begin{equation}
\label{eq:barxbartdagg}
\widetilde{X_{\bar t}^\dagger}=\sum_{N=0}^{N_{max}-1}\widetilde{X_{\bar t}^\dagger}(N)=(X_{\bar t}^\dagger)_L\mathcal{\hat Z}\breve 1_B^{(-1)}=\breve1_B(X_{\bar t}^\dagger)_L\mathcal{\hat Z}=\breve 1_B(X_{\bar t}^\dagger)_L\mathcal{\hat Z}\breve 1_B.
\end{equation}
In this way, we can obtain $(X_{\bar t'})_L$ and $(X_{\bar t}^\dagger)_L$ as infinite expansions.

Dealing with the terms up to $O(\Gamma^4)$, we have found that $\hat Z(N)=\mathcal{\hat Z}\hat 1_B(N)=\hat 1_B(N)\mathcal{\hat Z}$, $\widetilde{X_{\bar t'}}(N)=\widetilde{X_{\bar t'}}\hat 1_B(N+1)=\hat 1_B(N)\widetilde{X_{\bar t'}}$, and $\widetilde{X_{\bar t}^\dagger}(N)=\hat 1_B(N+1)\widetilde{X_{\bar t}^\dagger}=\widetilde{X_{\bar t}^\dagger}\hat 1_B(N)$.
Oppositely, assuming that these hold for any $N$, and substituting them into Eq. (\ref{eq:barxbtdagn}) and Eq. (\ref{eq:zn}), we can find the relational expressions for $\mathcal{\hat Z}$, $\widetilde{X_{\bar t'}}$, and $\widetilde{X_{\bar t}^\dagger}$, and solve these for each order of magnitude, then we obtain the same results. This result suggests that the $N$ dependency of these operators found up to $O(\Gamma^4)$ generally holds.

Applying the above results to Eqs. (\ref{eq:uoun}) and summing up $N$, we obtain
\begin{subequations}
\begin{equation}
\begin{array}{lll}
\widetilde{X_{t'}} &=&(X_{t'})_L\mathcal{\hat Z}\breve 1_B=\breve 1_B^{(-1)}(X_{t'})_L\mathcal{\hat Z}=\breve 1_B(X_{t'})_L\mathcal{\hat Z}\breve 1_B,
\\
(X_{t'})_L&=&(X_{t'})_D,
\end{array}
\end{equation}
\begin{equation}
\begin{array}{lll}
\widetilde{X_t^\dagger}&=&\breve 1_B(X_t^\dagger)_L\mathcal{\hat Z}=(X_t^\dagger )_L\mathcal{\hat Z}\breve 1_B^{(-1)}=\breve 1_B(X_t^\dagger )_L\mathcal{\hat Z}\breve 1_B,
\\
(X_t^\dagger)_L&=&(X_t^\dagger )_D+(X_t^\dagger )_{out}
\\
(X_t^\dagger )_{out}&=& -\displaystyle\frac 12\sum_{t_1t_2}\sum_{\bar t'_1}Y(tt_1t_2\bar t'_1)b_{t_1}^\dagger b_{t_2}^\dagger (X_{\bar t'_1})_L,
\end{array}
\end{equation}

\begin{equation}
\begin{array}{lll}
\label{eq:barbqn2}
\widetilde{B_q}&=&(B_q)_L\mathcal{\hat Z}\breve 1_B=\breve 1_B(B_q)_L\mathcal{\hat Z}=\breve 1_B(B_q)_L\mathcal{\hat Z}\breve 1_B,
\\
 (B_q) _L&=&(B_q)_D+(B_q)_{out},
\\
(B_q)_{out}&=&\displaystyle\sum_t\sum_{\bar t'}\Gamma_q^{\bar t't}b_t^\dagger(X_{\bar t'})_L.
\end{array}
\end{equation}
\end{subequations}

While, from $B_q=B_{\bar q}^\dagger$, $\widetilde{B_q}=\widetilde{B_{\bar q}}^\dagger$ and we find another expression for $\widetilde{B_q}$ as
\begin{equation}
\begin{array}{lll}
\widetilde{B_q}&=&\breve 1_B\mathcal{\hat Z}(B_{\bar q})_L^\dagger =\mathcal{\hat Z}(B_{\bar q})_L^\dagger\breve 1_B=\breve 1_B\mathcal{\hat Z}(B_{\bar q})_L^\dagger\breve 1_B,
\\
 (B_{\bar q}) _L^\dagger&=&(B_q)_D+(B_{\bar q})_{out}^\dagger,
\\
(B_{\bar q})_{out}^\dagger&=&\displaystyle\sum_t\sum_{\bar t}\Gamma_q^{t'\bar t}(X_{\bar t})_L^\dagger b_{t'},
\end{array}
\end{equation}
where we use $(B_q)_D=(B_{\bar q})_D^\dagger$.
Using two types of expressions for $\widetilde{B_q}$, we obtain
\begin{equation}
\label{eq:bqdzcom}
\breve 1_B[(B_q)_D, \mathcal{\hat Z}]\breve 1_B=\breve 1_B\{\mathcal{\hat Z}(B_{\bar q})_{out}^\dagger-(B_q)_{out}\mathcal{\hat Z}\}\breve 1_B.
\end{equation}

From Eq. (\ref{eq:bmopzubar}) and Eq. (\ref{eq:normopresp}), we can express the mapping operator as
\begin{equation}
\label{eq:mappingop}
U_\xi=\mathcal{\hat Z}^{\xi-\frac 12}\widetilde{U},
\end{equation}
and Eq. (\ref{eq:ximapzbar}) becomes as follows:
\begin{subequations}
\label{eq:ximapcalzbar}
\begin{equation}
\label{eq:ximapcalzbar1}
\vert \psi')_{\xi}= \mathcal{\hat Z}^{\xi-\frac 12}\widetilde{\vert\psi')},\qquad {}_{-\xi} (\psi\vert=\widetilde{(\psi\vert}\mathcal{\hat Z}^{-\xi-\frac 12},
\end{equation}
\begin{equation}
\label{eq:ximapcalzbar2}
(O_F)_{\xi}=\mathcal{\hat Z}^{\xi-\frac 12}\widetilde{O_F}\mathcal{\hat Z}^{-\xi-\frac 12},
\end{equation}
\end{subequations}
If $O_F$ is a phonon creation, a phonon annihilation, or a scattering operator, $\widetilde{O_F}=\breve 1_B(O_F)_L\mathcal{\hat Z}\breve 1_B$ holds. Therefore the mapped $O_F$ can be expressed as
\begin{equation}
\label{eq:ofxi}
(O_F)_\xi=\breve 1_B(O_F)_{B(\xi)}\breve 1_B;\quad (O_F)_{B(\xi)}=\mathcal{\hat Z}^{\xi-\frac 12}(O_F)_L\mathcal{\hat Z}^{-\xi+\frac 12},
\end{equation}
and 
\begin{equation}
{}_{-\xi}(\psi\vert(O_F)_\xi\vert\psi')_\xi={}_{-\xi}(\psi\vert(O_F)_{B(\xi)}\vert\psi')_\xi
\end{equation}
holds. Therefore we can regard $(O_F)_\xi$ as $(O_F)_{B(\xi)}$ in the physical subspace.
The boson expansions of $(O_F)_{B(\xi)}$ become infinite expansions for an arbitrary $\xi$ because those of $(O_F)_L$ become infinite expansions.

For $\xi\neq 0$,  the boson expansions become of the non-Hermitian type. In the case of $\xi=\frac12$,  $(O_F)_{\xi(\frac 12)}=(O_F)_L$ holds.

For $\xi=0$, the boson expansions become the Hermitian type and can be derived using
\begin{equation}
\begin{array}{lll}
(O_F)_{B(0)}&=&\mathcal{\hat Z}^{-\frac 12}(O_F)_L\mathcal{\hat Z}^{\frac 12}
\\
&=&(O_F)_L+\frac 12[(O_F)_L, \mathcal{\hat Y}]
-\frac 38\mathcal{\hat Y}[(O_F)_L, \mathcal{\hat Y}]-\frac 18[(O_F)_L, \mathcal{\hat Y}]\mathcal{\hat Y}+O(\Gamma^6).
\end{array}
\end{equation}

The boson expansions of the phonon creation and annihilation operators and the scattering operators are as follows:
\begin{subequations}
\label{eq:xtg4}
\begin{equation}
(X_{t'})_{B(0)}=(X_{t'})_{B(0) in}+(X_{t'})_{B(0) out},
\end{equation}
\begin{equation}
(X_{t'})_{B(0) in}=b_{t'}+(X_{t'})_{B(0) in}^{(2)}+(X_{t'})_{B(0) in}^{(4)}+O(\Gamma^5),
\end{equation}
\begin{equation}
(X_{t'})_{B(0) in}^{(2)}=-\frac14\sum_{t_1}\sum_{t'_1t'_2}Y(t't'_1t'_2t_1)b_{t_1}^\dagger b_{t'_1}b_{t'_2},
\end{equation}
\begin{equation}
\begin{array}{r}
(X_{t'})_{B(0) in}^{(4)}=-\displaystyle\frac{1}{32}\sum_{t_1}\sum_{t'_1t'_2}\sum_{tt''}Y(t'tt''t'_1)Y(t'_2tt''t'_1)b_{t_1}^\dagger b_{t'_1}b_{t'_2}
\\
+\displaystyle\frac 1{96}\sum_{t_1t_2}\sum_{t'_1t'_2t'_3}\sum_t\{2Y(t'_1t_1t't)Y(tt'_2t'_3t_2)
\\
-5Y(t't'_1t'_2t)Y(tt_1t_2t'_3)\}b_{t_1}^\dagger b_{t_2}^\dagger b_{t'_1}b_{t'_2}b_{t'_3},
\end{array}
\end{equation}
\begin{equation}
(X_{t'})_{B(0)out}=(X_{t'})_{B(0)out}^{(3)}+(X_{t'})_{B(0)out}^{(4)}+O(\Gamma^5)
\end{equation}
\begin{equation}
\begin{array}{r}
(X_{t'})_{B(0)out}^{(3)}=\displaystyle\frac 1{24}\sum_{t_1t_2}\sum_{t'_1t'_2t'_3}\sum_{\bar t}\{2Y(t'_1t_1t'\bar t)
Y(\bar tt'_2t'_3t_2)
\\
+Y(t'_1t_1t_2\bar t)Y(\bar tt'_2t'_3t')\}b_{t_1}^\dagger b_{t_2}^\dagger b_{t'_1}b_{t'_2}b_{t'_3},
\end{array}
\end{equation}
\begin{equation}
\begin{array}{r}
(X_{t'})_{B(0)out}^{(4)}=-\displaystyle\frac 1{16}\sum_{t_1t_2t_3}\sum_{t'_1t'_2t'_3t'_4}\sum_{\bar t\bar t'}Y(t'_1t't_1\bar t)Y(\bar tt'_2t'_3\bar t')Y(\bar t't_2t_3t'_4)
\\
b_{t_1}^\dagger b_{t_2}^\dagger b_{t_3}^\dagger b_{t'_1}b_{t'_2}b_{t'_3}b_{t'_4}.
\end{array}
\end{equation}
\end{subequations}
\begin{subequations}
\label{eq:xtbarg4}
\begin{equation}
(X_{\bar t'})_{B(0)}=(X_{\bar t'})_{B(0)}^{(2)}+(X_{\bar t'})_{B(0)}^{(3)}+(X_{\bar t'})_{B(0)}^{(4)}+O(\Gamma^5) ,
\end{equation}
\begin{equation}
(X_{\bar t'})_{B(0)}^{(2)}=(X_{\bar t'})_L^{(2)},\quad (X_{\bar t'})_{B(0)}^{(3)}=(X_{\bar t'})_L^{(3)},
\end{equation}
\begin{equation}
(X_{\bar t'})_{B(0)}^{(4)}=(X_{\bar t'})_L^{(4)}-\frac 12[ \mathcal{\hat Z}^{(2), }(X_{\bar t'})_L^{(2)}].
\end{equation}
\end{subequations}
\begin{subequations}
\label{eq:bqg4}
\begin{equation}
\begin{array}{lll}
(B_q)_{B(0)}&=&(B_q)_L+\frac 12\mathcal{\hat Z}\{(B_{\bar q})_{out}{}^\dagger-(B_q)_{out}\}+O(\Gamma^5),
\\
&=&(B_q)_{B(0)in}+(B_q)_{B(0)out},
\end{array}
\end{equation}
\begin{equation}
(B_q)_{B(0)in}=(B_q)_D,
\end{equation}
\begin{equation}
(B_q)_{B(0)out}=(B_q)_{B(0)out}^{(2)}+(B_q)_{B(0)out}^{(3)}+(B_q)_{B(0)out}^{(4)}+O(\Gamma^5),
\end{equation}
\begin{equation}
(B_q)_{B(0)out}^{(k)}=\frac 12\{(B_q)_{out}^{(k)}+(B_{\bar q})_{out}^{(k)}{}^\dagger\}\quad (k=2,3),
\end{equation}
\begin{equation}
(B_q)_{B(0)out}^{(4)}=\frac 12\{(B_q)_{out}^{(4)}+(B_{\bar q})_{out}^{(4)}{}^\dagger\}+\frac 12\mathcal{\hat Z}^{(2)}\{(B_{\bar q})_{out}^{(2)}{}^\dagger-(B_q)_{out}^{(2)}\},
\end{equation}
\begin{equation}
\begin{array}{r}
\frac 12\mathcal{\hat Z}^{(2)}\{(B_{\bar q})_{out}^{(2)}{}^\dagger-(B_q)_{out}^{(2)}\}=\displaystyle\frac 18\sum_{t_1t_2}\sum_{t'_1t'_2}\sum_{tt'}\sum_{\bar t}\{\Gamma_q^{t'_1\bar t}Y(\bar ttt't'_2)-\Gamma_q^{\bar tt}Y(\bar tt'_1t'_2t')\}Y(tt_1t_2t')
\\
b_{t_1}^\dagger b_{t_2}^\dagger b_{t'_1}b_{t'_2}
\\
+\displaystyle\frac 18\sum_{t_1t_2t_3}\sum_{t'_1t'_2t'_3}\sum_{t}\sum_{\bar t}\{2\Gamma_q^{t'_1\bar t}Y(\bar tt_1tt'_2)-\Gamma_q^{\bar tt}Y(\bar tt'_1t'_2t_1)-\Gamma_q^{\bar tt_1}
Y(\bar tt'_1t'_2t)\}
\\
Y(tt_1t_2t'_3)b_{t_1}^\dagger b_{t_2}^\dagger b_{t_3}^\dagger b_{t'_1}b_{t'_2}b_{t'_3}
\\
+\displaystyle\frac1{16}\sum_{t_1t_2t_3t_4}\sum_{t'_1t'_2t'_3t'_4}\sum_{\bar t}\{\Gamma_q^{t'_1\bar t}Y(\bar tt_1t_2t'_2)-\Gamma_q^{\bar tt_1}Y(\bar tt'_1t'_2t_2)\}
Y(t'_4t_3t_4t'_3)
\\
b_{t_1}^\dagger b_{t_2}^\dagger b_{t_3}^\dagger b_{t_4}^\dagger b_{t'_1}b_{t'_2}b_{t'_3}b_{t'_4}.
\end{array}
\end{equation}
\end{subequations}
Here, we use Eq. (\ref{eq:bqdzcom}) to find $(B_q)_{B(0)}$. From Eq. (\ref{eq:barbqn2}), $(B_q)_{out}^{(k)}=\displaystyle\sum_t\sum_{\bar t'}\Gamma_q^{\bar t't}b_t^\dagger(X_{\bar t'})_L^{(k)}$.

Finally, we deal with the product of operators.
Let $O_F$ and $O'_F$ be the phonon creation, annihilation operators, or scattering operators, respectively, we can derive the boson expansions of their product as 
\begin{equation}
\label{eq:ofof'exp}
(O_FO'_F)_{B(\xi)}=\mathcal{\hat Z}^{\xi-\frac 12}\widetilde{O_FO'_F}\mathcal{\hat Z}^{-\xi-\frac 12}.
\end{equation}
If $\widetilde{O_FO'_F}=\breve 1_B(O_F)_L\breve 1_B(O'_F)_L\mathcal{\hat Z}\breve 1_B$ holds,  we obtain
\begin{equation}
\label{eq:ofof'exppr}
(O_FO'_F)_{B(\xi)}=(O_F)_{B(\xi)}(O'_F)_{B(\xi)},
\end{equation}
and if Eq. (\ref{eq:ximapzbar22aprox}) holds,
\begin{equation}
\label{eq:ofof'expprapp}
(O_FO'_F)_{B(\xi)}\approx (O_F)_{B(\xi)}(O'_F)_{B(\xi)}.
\end{equation}
In the case that Eq. (\ref{eq:ofof'exppr}) and Eq. (\ref{eq:ofof'expprapp}) hold, it is sufficient to 
obtain only the boson expansions of the basic fermion pair operators.
Conventional practical boson expansion methods have used, as a matter of course, the approximation of Eq. (\ref{eq:ofof'expprapp}).
Eq. (\ref{eq:ofof'exp}) makes it possible to judge whether this approximation is good or bad.
We present $\widetilde{O_FO'_F}$ in the appendix.

We finally point out that $\bar t$-sum does not need to sum all $\bar t$ in the following cases.
\begin{equation}
\label{eq:doublecomtau}
[ [X_{\tau_1}, X_{\tau_2}^\dagger], X_{\tau_3}^\dagger] \approx -\sum_{\tau'}Y(\tau_1, \tau_2, \tau_3, \tau ')X_{\tau'}^\dagger,
\end{equation}
are satisfied for the phonon excitation modes $\{\tau\}$, and $\{\tau\}$ contains $\{t\}$ and is set up such that the small parameter expansion holds, then $\bar t$ not contained in $\{\tau\}$ can be neglected in $\bar t$-sum.
An example is a case where $\{\tau\}$ contains a sufficient variety of phonon excitation modes and
\begin{equation}
\sum_\tau\psi_\tau(\alpha\beta)\psi_\tau(\alpha'\beta')\approx\delta_{\alpha\alpha'}\delta_{\beta\beta'}-\delta_{ \alpha\beta'}\delta_{\beta\alpha'}
\end{equation}
are satisfied.
In this case,
\begin{equation}
a_\alpha^\dagger a_\beta^\dagger\approx \sum_\tau\psi_\tau(\alpha\beta)
\end{equation}
are satisfied, therefore,
\begin{equation}
X_{\bar\tau}^\dagger\approx 0
\end{equation}
are satisfied, and 
Eq. (\ref{eq:doublecomtau}) are derived.
In this case, however, $\{\tau\}$ cannot be regarded as $\{t\}$ because $\{\tau\}$ contains so sufficient variety of phonon excitation modes that $\hat Z$ includes zero eigenvalues.

\subsection{Boson expansions in the case where the double commutators of the phonon operators are closed}
\label{dcommphononcl}
In this section, we treat the boson expansions in the case where the double commutators of Eq. (\ref{eq:doublecom}) are closed in $\{t\}$.

If $Y(t'_1t\bar tt'_2)=0$, then the double commutators of Eq. (\ref{eq:doublecom}) are closed in $\{t\}$.
For further analysis, we denote more concretely $\mathbf{W}(2)$ and $\mathbf{Z'}(2)$ in Eq. (\ref{eq:zwwtz'}) as follows,
\begin{subequations}
\begin{equation}
\label{eq:w12}
\mathbf{W}(2)=\left (\mathbf{W}^{(1)}(2)\  \mathbf{W}^{(2)}(2)\right ),
\end{equation}
\begin{equation}
\label{eq:z'123}
\mathbf{Z'}(N)=\left (
\begin{array}{@{\,}cc@{\,}}
\mathbf{Z}^{(1)}(2)&\mathbf{Z'}^{(3)}(2)
\\
\mathbf{Z'}^{(3)}(2)^T&\mathbf{Z'}^{(2)}(2)
\end{array}
\right ),
\end{equation}
\end{subequations}
where $\mathbf{W}^{(1)}(2)$ is what becomes a zero matrix when $Y(t'_1t\bar tt'_2)=0$. 
Substituting this into Eq. (\ref{relzwz'2}), then $\mathbf{W}^{(2)}(2)\mathbf{Z'}^{(3)}(2)^T$ becomes a zero matrix, and we obtain
\begin{equation}
\label{eq:spimp}
\sum_{\bar t\bar t'}Y(t'_1\bar t\bar t't'_2)Y(t_1\bar t\bar t'\bar t_1)=0,
\end{equation}
which indicates that Eqs. (\ref{eq:spestres}) do not hold.
It also indicates that if $Y(t'_1t\bar tt'_2)=0$, then $Y(t'_1\bar t_1\bar t_2t'_2)=0$ should be satisfied. $\mathbf{W}(2)=\mathbf{0}(2)$ holds, and $\vert t_1t_2\rangle$ and $\vert\bar t\mu\rangle$ become orthogonal.
Therefore, if the double commutators of Eq. (\ref{eq:doublecom}) are closed in $\{t\}$, the boson expansions are not obtained as the small parameter expansion.

Starting with Eqs. (\ref{eq:uoun}) and applying $Y(t'_1\bar t\mu t'_2)=0$ to them, we derive $\widetilde X_{\bar t}^\dagger (N)=0 $, from which we obtain $\widetilde X_{\bar t}=0$ and $\widetilde X_{\bar t}^\dagger=0$.
Therefore, $(X_{\bar t'})_\xi=0$ and $(X_{\bar t}^\dagger )_\xi=0$ hold.
Inversely, if $(X_{\bar t'})_\xi=0$ and $(X_{\bar t}^\dagger )_\xi=0$ hold, then $Y(t'_1\bar t\mu t'_2)=0$ should hold, that is $\vert t'_1t'_2\rangle$ and $\vert\bar t\mu\rangle$ should be orthogonal
because
\begin{equation}
Y(t'_1\bar t\mu t'_2)=((t'_1t'_2\vert\bar t\mu))-\langle\langle t'_1t'_2\vert\bar t\mu\rangle\rangle,
\end{equation}
and
\begin{equation}
\langle\langle t'_1t'_2\vert\bar t\mu\rangle\rangle=\langle 0\vert X_{t'_1}X_{t'_2}X_{\bar t}^\dagger X_{\mu}\vert 0\rangle
=(0\vert (X_{t'_1})_\xi (X_{t'_2})_\xi (X_{\bar t}^\dagger)_\xi (X_{\mu})_\xi\vert 0)
\end{equation}
hold.
It is a necessary and sufficient condition for $(X_{\bar t'})_\xi=0$ and $(X_{\bar t}^\dagger )_\xi=0$ that $\vert t'_1t'_2\rangle$ and $\vert\bar t\mu\rangle$ are orthogonal.
We also obtain $\widetilde X_{t'}=(X_{t'})_D\hat Z$, $\widetilde X_t^\dagger =(X_t^\dagger)_D\hat Z$, and $\widetilde B_q=(B_q)_D\hat Z$.
Therefore,
\begin{equation}
\label{eq:ofxi1}
(O_F)_\xi=\hat Z^{\xi-\frac 12}(O_F)_D\hat Z^{-\xi+\frac 12},
\end{equation}
where $O_F$ is $X_{t'} $, $ X_t^\dagger$, or $B_q$.

From Eq, (\ref{eq:scop2}) and Eq. (\ref{eq:barbqn}), $[(B_q)_D, \hat Z(N)]=0$ holds, then
\begin{equation}
\label{eq:bqzncom}
[(B_q)_D, \hat Z]=0.
\end{equation}
Hence
\begin{equation}
\label{eq:bqxibqd}
(B_q)_\xi= (B_q)_D\hat T_B=\hat T_B(B_q)_D=\hat T_B(B_q)_D\hat T_B
\end{equation}
holds for any $\xi$.

For $O_F$ and $O'_F$ being the phonon operators or the scattering operators, respectively, we obtain
\begin{equation}
\label{eq:ofxi2}
(O_FO'_F)_\xi=\hat Z^{\xi-\frac 12}(O_F)_D(O'_F)_D\hat Z^{-\xi+\frac 12},
\end{equation}
In addition, the following,
\begin{equation}
\label{eq:ofxi2ofof}
(O_FO'_F)_\xi=(O_F)_\xi(O'_F)_\xi,
\end{equation}
is satisfied if $\hat T_B$ becomes $\breve 1_B$ and $O_FO'_F$ is normal ordered
because $\hat Z^{\xi-\frac 12}\hat Z^{-\xi+\frac 12}=\breve 1_B$ and $\breve 1_B(X_t^\dagger)_D=\breve 1_B(X_t^\dagger)_D\breve 1_B$, $(X_{t'})_D\breve 1_B=\breve 1_B(X_{t'})\breve 1_B$, and Eq. (\ref{eq:bqzncom}) are satisfied.

Eq. (\ref{eq:zn}) becomes
\begin{equation}
\label{eq:zncaapprox}
\hat Z(N)=\frac 1N\sum_t(X_t^\dagger)_D\hat Z(N-1)b_t\quad (N\ge 2).
\end{equation}

The solution of Eq. (\ref{eq:zncaapprox}) should be given by Eq. (\ref{eq:znnons1}).
While Eq. (\ref{eq:zncaapprox}) can be solved directly in the case that $\mathbf{Z}(2)$ has no zero eigenvalues.
In this case, $Y(t'_1t_1t_2t'_2)=-2((t'_1t'_2\vert t_1t_2))$ holds.
Therefore,
\begin{equation}
\label{eq:xtdcag}
(X_t^\dagger)_D=b_t^\dagger (2\hat N_B+1),
\end{equation}
from which we obtain
\begin{equation}
\label{eq:zncaapprox1b}
\hat Z(N)=\frac {(2N-1)}N\sum_tb_t^\dagger \hat Z(N-1)b_t.
\end{equation}
$\hat Z(2)=3\hat1_B(2)$, and if $\hat Z(N-1)=(2N-3)!!\hat 1_B(N-1)$, then $\hat Z(N-1)=(2N-1)!!\hat 1_B(N)$. 
These match Eq. (\ref{eq:znnons1}) in the case that $\hat T_B(N)=\hat 1_B(N)$ holds.

Eq. (\ref{eq:zncaapprox}) can be also solved formally as
\begin{equation}
\label{eq:formsolzn}
\hat Z(N)
=\frac1{N!}\sum_{t_1\cdots t_N}(X_{t_1}^\dagger)_D\cdots(X_{t_N}^\dagger)_D\vert 0)(0\vert b_{t_1}\cdots b_{t_N}.
\end{equation}

From Eq. (\ref{eq:formsolzn}), we find the relation,
\begin{equation}
\label{eq:zbxtdz}
\hat Z(N)b_t^\dagger=(X_t^\dagger )_D\hat Z(N-1).
\end{equation}
While, from Eq. (\ref{eq:znnons1}) and Eq. (\ref{eq:zncaapprox}), we obtain
\begin{equation}
\label{eq:tbxtdrel}
(2N-1)\hat T_B(N)b_t^\dagger =(X_t^\dagger)_D\hat T_B(N-1).
\end{equation}

The mapped operators are given as follows,
\begin{subequations}
\label{eq:ofbxinons}
\begin{equation}
\label{eq:ofbxinons1}
(O_F)_\xi=\hat T_B(O_F)_{B(\xi)}\hat T_B,
\end{equation}
\begin{equation}
\label{eq:ofbxinons2}
(O_F)_{B(\xi)}=\left\{(2\hat N_B-1)!!\right\}^{\xi-\frac 12}(O_F)_D\left\{(2\hat N_B-1)!!\right\}^{-\xi+\frac 12}
\end{equation}
\end{subequations}
From Eq. (\ref{eq:ofxi2}), we obtain
\begin{equation}
\label{eq:of2bxinons}
(O_FO'_F)_{B(\xi)}=\left\{(2\hat N_B-1)!!\right\}^{\xi-\frac 12}(O_F)_D(O'_F)_D\left\{(2\hat N_B-1)!!\right\}^{-\xi+\frac 12}.
\end{equation}
The difference due to $\xi$ is renormalized to the boson excitation number, and the remaining boson expansions are the same as those of the DBET, which are finite.
Therefore,  we can substantially treat all types of boson expansions as finite expansions when the concerning states are the physical states that are eigenstates of the boson number operator.
If $\xi=0$, we obtain finite boson expansions of the Hermitian type.

In the case that $O_F$ preserves the number of quasi-particles, then $(O_F)_D$ preserves the number of bosons, and the norm operator parts cancel out completely. As a result, we obtain finite  boson expansions for any $\xi$ such as
\begin{equation}
(O_F)_{B(\xi)}=(O_F)_D,
\end{equation}
from which we also derive
\begin{equation}
(O_F)_{B(0)}=(O_F)_{B(\xi)}.
\end{equation}
Even if $O^{(1)}_F$ and $O^{(2)}_F$ do not necessarily preserve the quasi-particle number, respectively, if $O_F=O^{(1)}_FO^{(2)}_F$ preserves the quasi-particle number,Eq. (\ref{eq:of2bxinons}) enables us to derive
\begin{equation}
(O^{(1)}_FO^{(2)}_F)_{B(0)}=(O^{(1)}_FO^{(2)}_F)_{B(\xi)}=(O^{(1)}_F)_D(O^{(2)}_F)_D.
\end{equation}
Hence $(O_F)_D$ and $(O^{(1)}_F)_D(O^{(2)}_F)_D$ are regarded as a finite boson expansion of the Hermitian type.

For $\xi=\frac 12$, the norm operator does not appear in the mapped fermion operators, and we can obtain the boson expansions as follows,
\begin{equation}
\label{eq:oaaprox12}
(O_F)_{B(\frac 12)}=(O_F)_D,
\end{equation}
\begin{equation}
\label{eq:oo'caapprox12}
(O_FO'_F)_{B(\frac 12)}=(O_F)_{B(\frac 12)}(O'_F)_{B(\frac 12)}=(O_F)_D(O'_F)_D.
\end{equation}
We obtain the finite expansions of DBET. 
From Eqs. (\ref{eq:nhhr}), it is straitforward to proof that Hermitian treatment\cite{Ta01} holds exactly for the eigenvector of $\hat Z$, $\vert N; a)$.
Therefore, when $\hat T_B=\hat 1_B$ holds, it is applied precisely for the ideal boson state vector, $\vert N: t)$.

For $\xi=0$, the boson mapping becomes of the Hermite type.
We can obtain mapped operators as follows,
\begin{subequations}
\label{eq:xtbqb0}
\begin{equation}
\label{eq:xtdhptb}
\begin{array}{lll}
(X_t^\dagger)_{B(0)}&=&\left\{(2\hat N_B-1)!!\right\}^{-\frac 12}(X_t^\dagger )_D\left\{(2\hat N_B-1)!!\right\}^{\frac 12},
\\
&=&(X_t^\dagger )_D\displaystyle\frac1{\sqrt{1+2\hat N_B}}
=b_t^\dagger\sqrt{1+2\hat N_B},
\end{array}
\end{equation}
\begin{equation}
(B_q)_{B(0)}=\sum_{tt'}\Gamma_q^{t't}b_t^\dagger b_{t'}.
\end{equation}
\end{subequations}
Here we use the relation,
\begin{equation}
\hat T_Bb_t^\dagger (2\hat N_B+1)=(X_t^\dagger )_D\hat T_B,
\end{equation}
obtained from Eq. (\ref{eq:tbxtdrel}) for the derivation of $(X_t^\dagger)_{B(0)}$.
The scattering operators are expressed as finite expansions in the physical subspace.
The phonon operators do not become of the small parameter expansion whose zeroth-order approximation becomes the boson approximation.
The boson approximation holds only when the phonon excitation number does not exceed one.

\subsection{On the role of the norm operator}
In this subsection,  based on the results obtained so far, we summarize the role the norm operator plays in the boson expansion method.
What is important is the relation between the norm operator consisting of all kinds of phonons and the norm operator constituting the boson mapping operator.

What kinds of boson expansions are derived is determined by the structure of the norm operator when all modes are adopted, $\hat Z^{(A)}$.
The structure is determined by the introduced single-particle states, $\{\alpha\}$, and the amplitudes of the Tam-Dancoff phonons, $\psi_\mu(\alpha\beta)$ .
$\hat Z^{(A)}$ is composed of the norm operator $\hat Z$, which is used for mapping, and other operators as
\begin{subequations}
\begin{equation}
\hat Z^{(A)}=\hat Z+\hat W+\hat W^\dagger+\hat Z,'
\end{equation}
where
\begin{equation}
\hat Z=\breve 1_B\hat Z^{(A)}\breve 1_B,
\quad\hat W=\breve 1_B\hat Z^{(A)}(\breve 1_B^{(A)}-\breve 1_B),
\quad\hat Z'=(\breve 1_B^{(A)}-\breve 1_B)\hat Z^{(A)}(\breve 1_B^{(A)}-\breve 1_B),
\end{equation}
\end{subequations}
The condition Eq. (\ref{eq:zacondi}) is imposed on $\hat Z^{(A)}$, which is regardless of how to take $\psi_\mu(\alpha\beta)$. 
$\hat Z$, $\hat W$, and $\hat Z'$ are determined so as to satisfy the condition Eq. (\ref{eq:zacondi}).

The double commutation relations of the phonon operators that constitute $\hat Z$ are generally not closed among them.
$\hat Z$ must have eigenvalues nearly to 1 for the small parameter expansion, which also allows the use of ideal boson state vectors as physical.
It is possible to directly check whether this condition is satisfied because $\hat Z$ is specifically obtained by the boson expansion assuming the small parameter expansion.
Only this condition is not, however, a sufficient condition for realizing the small parameter expansion.
Including not only $\hat Z$ but also $\hat W$ and $\hat Z'$ gives the necessary and sufficient condition.
In this case, it is not allowed to treat $\hat W$ as a zero operator, that is, to assume that the double commutation relation of the phonon operators constituting $\hat Z$ is closed because the small parameter expansion does not hold.

In the case that $\hat W$ can be regarded as zero, the boson expansions can be substantially treated as finite expansions.
The realization of this type of practical boson expansion is more difficult because $\{t\}$ should selected so that the ideal boson state vectors become physical and the dynamics are reflected enough, as with the small parameter expansion, under the above condition.

\section{Comments on the conventional methods}
\label{comp}
Conventional practical boson expansion methods, without exception, discard the phonon excitation modes that are not adopted for the boson excitation modes.
We call this procedure the non-adopted modes discarding (NAMD).

Since NAMD closes double commutators of phonon operators within the adopted modes for those of bosons, it is incompatible with a small expansion.
The incompatibility between NAMD and the small parameter expansion has not been considered in formulating the conventional practical boson expansion methods.

In the case that NADM is precisely applicable, DBET is formulated exactly, which does not mean that DBET necessarily has its exceptional superiority because we can substantially obtain finite expansions of the Hermite type by treating the boson number operator parts appropriately.

In the case that the small parameter expansion is applicable, all the boson expansions become infinite ones and include the terms neglected by NAMD.
Applying NADM to Eq. (\ref{eq:xtg4}), Eq. (\ref{eq:xtbarg4}), and Eq. (\ref{eq:bqg4}), it is found that the remaining terms up to $O(\Gamma^2)$ coincide with those obtained by NOLCEXPT.
The order of magnitude of the neglected terms is also $O(\Gamma^2)$, respectively.
NOLCEXPT obtains, with NADM, the terms only up to $O(\Gamma^2)$.
On the other hand, the finite boson expansions of DBET are obtained from$(O_F)_{B(\frac 12)}$ by applying NADM.
The order of magnitude of the neglected terms by NADM is also $O(\Gamma^2)$ in the same order as the smallest one of the non-neglected terms by NADM.
In both NOLCEXPT and DBET, NADM neglects terms of the order of magnitude that should be adopted,
which indicates that NAMD can not be used as a proper approximation under the small parameter expansion.

The investigation so far makes it clear that the comment of NOLCEXPT \cite{KT83, SK88} on NAMD is incorrect.
NOLCEXPT claims that the scattering operators are expressed as finite expansions.
It is realized only when NADM is precisely or well approximately applied and not when the small parameter expansion is realized.
Eqs. (\ref{eq:xtbqb0}) indicate that it is impossible to express the phonon operators as infinite normal-ordered small parameter expansions although the scattering operators become finite.
NOLCEXPT has failed to refute Marshalek's claim \cite{Ma80a, Ma80b} that KT-1 \cite{KT72} and KT-2 \cite{KT76} are chimerical boson expansions.

As already mentioned, Hermitian treatment becomes exact when NAMD becomes exact. 
On the other hand, in the case that the small parameter expansion is applicable, 
Hermitian treatment becomes an approximation, and it can be generally evaluated using the norm operator by following the method of \cite{KTM99}.
It is concluded that Hermitian treatment holds as far as it is possible to neglect terms of $O(\Gamma^4)$.

Next, we comment on the problems related to a modified Marumori boson expansion method \cite{HJJ76, MTS81}.
The modified Marumori boson expansion method concludes that NADM is good using a norm of a multi-phonon state vector despite the small parameter expansion being available \cite{HJJ76}.
The reason why the conclusion is derived incorrectly is as follows:
The norm of the multi-phonon state vector is obtained from  $\hat Z(N)$.
Since the neglected terms by NADM do not appear up to $O(\Gamma^3)$ in $\hat Z(N)$, 
it is impossible to evaluate whether NADM is a good approximation by investigating $\hat Z(N)$ only up to $O(\Gamma^2)$ with the small parameter expansion.
For explanation, we adopt a case where $\{t\}$ consists of only one type of excitation mode $c$.
We define $\vert N\rangle$ and $\vert N)$ as
\begin{equation}
\label{eq:Nc}
\vert N\rangle =\vert\overbrace{c\cdots c}^N\rangle,\quad \vert N)=\vert\overbrace{c\dots c}^N).
\end{equation}
They sattisfy $\langle N\vert N\rangle=(N\vert\hat Z(N)\vert N)$. 
Assuming the small parameter expansion, and setting $\langle 2\vert 2\rangle=1-\varepsilon$, then $\varepsilon=\frac 12Y(cccc)\sim O(\Gamma^2)$.
Expressing $\langle N\vert N\rangle$ derived by the small parameter expansion up to $O(\Gamma^2)$ as ${\mathcal N}^{(2)}(N)$, we obtain $\mathcal{N}^{(2)}(N)=1-N(N-1)\frac\varepsilon 2+O(\Gamma^3)$.
On the other hand, Eq. (\ref{eq:zncaapprox}) enables us to obtain $\hat Z(N)$ with the sum of all terms without the neglected ones by NADM.
Expressing $\langle N\vert N\rangle$ thus obtained as ${\mathcal N}^{(all)}(N)$, 
\begin{equation}
\label{eq:recc}
{\mathcal N}^{(all)}(N)={\mathcal N}^{(all)}(N-1)\left (1+(\langle 2\vert 2\rangle-1)(N-1)\right ),
\end{equation} 
holds, and we obtain ${\mathcal N}^{(all)}(N)=(1-(N-1)\varepsilon)(1-(N-2)\varepsilon)\cdots(1-\varepsilon)$.
As $N$ becomes large, the difference between both becomes prominent, which is, however, only $2\varepsilon^2\sim O(\Gamma^4)$ for $N=3$. 
It indicates that the small parameter expansion is well applied for this case
and that both coincide well up to $O(\Gamma^2)$ with the exact one.
Therefore it is impossible to judge whether NADM is good or not by the comparison of ${\mathcal N}^{(all)}(3)$ with the exact one.
It is a wrong conclusion that NADM holds well for $\varepsilon \approx 0.1$ \cite{HJJ76}, where the small parameter expansion becomes possible. $\varepsilon$ should become approximately -2 for NADM to become good. 
In addition, it is not the strong but the weak effect of the Pauli exclusion principle that makes the small parameter expansion possible.
The comment on the convergence of the modified Marumori boson expansions is mistaken.

Conventional practical boson expansion methods restrict, for mapping, only the sort of phonon excitation modes and not the number of those.
$\hat Z(N)$ necessarily has zero eigenvalues for large phonon excitation numbers even if restricting the sorts of modes, 
which makes ideal boson state vectors unphysical and the small parameter expansion impossible.
We should restrict the phonon excitation number beforehand.
Nevertheless, NOLCEXPT does not restrict the phonon excitation number.
Instead, it treats, without a clear basis, all the norm matrices of the multi-phonon state as having no zero eigenvalues.
Nevertheless, it gives the correct expansion terms up to $O(\Gamma^2)$ under NADM.
The restriction of the phonon excitation number beforehand gives a clear reason for it
because we obtain 
\begin{equation}
\label{eq:nmaxinfty}
\lim_{N_{max}\rightarrow \infty}(O_F)_\xi=(O_F)_{B(\xi)}
\end{equation}
from  Eq. (\ref{eq:ofxi}).
It indicates that we can obtain the correct results of the small parameter expansion without limiting the phonon excitation number beforehand.
Afterward, we should limit the boson excitation number.

As for BFEXP, by replacing the collective modes $\{c\}$ and the non-collective modes $\{n\}$ to $\{t\}$ and $\{\bar t\}$, respectively, and suppressing the fermion excitations, we can obtain the boson expansions from BFEXP  \cite{TM90, TM91}. 
The Hermitian-type boson expansions obtained thus agree with those obtained from BFEXP by adopting the proper transformation \cite{TKM94}.
Further comparison requires the derivation of higher-order terms in BFEXP.

\section{Comments on the application to the collective motions of nuclei}
\label{appl}
Here, we comment on the application of NOM to the collective motions of nuclei.

Although boson expansion methods have been applied to elucidate large-amplitude collective motions such as a shape transition of Sm isotopes \cite{KT76, TTT87, SK91}, there is room for examination as to whether they have been appropriately applied.
In order to conclude whether the boson expansion methods are of successful microscopic theory for elucidating the collective motion of nuclei, we must confirm whether the boson expansion methods are not applied beyond the condition to hold.
NOM can offer the necessary conditions so that the small parameter expansion holds, which the conventional methods can not.

NADM does not hold for the small parameter expansion.
The failure of NADM becomes a necessary condition for the availability of the small parameter expansion.
NOM takes up the contribution of the phonon excitation modes neglected by NADM.
The phonon operator with the collective excitation mode is sought as a superposition of spin-2 quasi-particle pairs.
We investigate the contribution of the phonon excitation modes neglected in NADM by taking up first those having spin-2 the same as the collective excitation mode as a trial, then gradually expanding to spin-4 and beyond.

Whether or not the ideal boson state vectors are physical can be determined by obtaining the eigenvalues of the norm operator.
The norm operator is given by $\mathcal{\hat Z}$ in the case that the small parameter expansion holds.
The eigenvalues of $\mathcal{\hat Z}$ must not deviate so much from 1, which becomes the necessary condition both the small parameter expansion holds and the ideal boson state vectors become physical.
Its eigenvalues are obtained using the boson expansion of $\mathcal{\hat Z}$.
When some non-collective excitation modes are introduced as boson excitations in addition to collective excitation modes, the energy eigenvalues are obtained by diagonalizing the Hamiltonian with the different excitation numbers for collective and non-collective excitation modes, respectively.
The excitation number is 1 or less for the non-collective modes, while a more excitation number is admitted for the collective modes.
Dynamical point of view would justify it.
This is, however, an inappropriate way for the norm operator.
Boson expansion methods treat all the phonon excitation modes mapped as boson excitations equally without distinction of their types, collective or non-collective.
Therefore, the effect of the Pauli exclusion principle can not be evaluated correctly by limiting the exciting number of specific modes.
The eigenvalues of the norm operator must be obtained by treating all the excitation modes adopted as bosons equally.
Since the norm operator commutes with the total boson number operator, it is sufficient to find eigenvalues only in the case of the maximum boson excitation number.
Perturbation theory is a simple method for obtaining eigenvalues.
Divide the norm operator $\mathcal{\hat Z}$ into $\breve 1_B+\mathcal{\hat Y}$ and regard $\mathcal{\hat Y}$ as a perturbation and find the eigenvalues of $\mathcal{\hat Z}$. The diagonal components of $\mathcal{\hat Z}$ are the eigenvalues obtained by first-order approximation correction.

In some cases, RPA-type correlations are incorporated \cite{KT76, SK88, SK91}.
New bosons are introduced by transforming the original ones.
This transformation should be applied to $(O_F)_\xi$, but in fact it has been applied to $(O_F)_{B(\xi)}$.
Eq. (\ref{eq:ofxi}) indicates that this procedure is justified only when the transformation does not change $\breve 1_B$. 
It is only when the correlation is weak enough that $\breve 1_B$ with proper $N_{max}$ for boson expansions becomes approximately unchanged.
On the other hand, if $N_{max}$ is large enough, $\breve 1_B$ can be regarded as $\hat 1_B$, the unit operator of the boson space, and bears no change under the transformation.
Therefore, this method of incorporating the RPA-type correlation needs the boson expansions derived by the mapping operator that allows an infinite number of phonon excitations in order to become applicable regardless of the strength of the correlation. 
As mentioned after Eq. (\ref{eq:nmaxinfty}), the boson expansions obtained by limiting the phonon excitation number with $N_{max}$ as the maximum beforehand can be derived from those obtained without such a limitation.
The phonon creation and annihilation operators correspond one-to-one to the boson creation and annihilation operators with the same excitation modes, respectively.
Therefore, limiting the boson excitation number by $N_{max}$ as the maximum afterward is the same as restricting the phonon excitation number by $N_{max}$ as the maximum beforehand.
In this way, we can limit the phonon excitation number properly by limiting the boson excitation number afterward.
However, the transformation to the new bosons breaks such a one-to-one correspondence.
Since the vacuum of the new bosons composes of the superposition of the original boson state vectors with excitation numbers from zero to infinity, 
merely setting the maximum excitation number of the new bosons can not limit the excitation numbers of the original bosons.
Therefore, even if we limit the number of excitations of the new bosons, the conditions for the small parameter expansion might not be satisfied when the introduced correlation becomes too strong.
Conventional methods, without the limitation for the phonon excitation number in the mapping operator, do not consider this point.
We should obtain the eigenvalues of $\mathcal{\hat Z}$ to investigate.
$\mathcal{\hat Z}$ does not commute with the total number operator of these new bosons.
Therefore, the eigenvalues of $\mathcal{\hat Z}$ should be obtained in the new boson space for all excitation numbers up to the maximum treating all boson excitation modes equally.
We can mention perturbation theory as a practical method.

Unlike the case where small parameter expansion can be applied,  it becomes difficult for the finite boson expansion to determine whether the ideal boson state vectors are physical.
The norm operator given by Eq. (\ref{eq:znnons2}) includes the projection operator of the physical space, $\hat T_B$.
Since $\hat T_B$ is not tractable, we should diagonalize $\hat Z$ itself and investigate whether its eigenvalues become zero in order to determine whether the ideal boson state vectors are physical.
Besides, we should establish a method to find such an orthogonal transformation of Eq. (\ref{eq:phononopc}) that makes NADM well satisfied and makes the dynamics of the system reflected enough.

The elucidation of the interacting boson model (IBM) by microscopic theory is an important subject, and various attempts have been made \cite{KM91}.
Since IBM conserves the total boson number, the finite boson expansion can be applied as a Hermite type.
The above issues, however, should be solved to realize and apply the finite boson expansion to it.

\section{Summary}
\label{sum}

We have proposed a new boson expansion theory, the norm operator method, where the norm operator plays a crucial role.

The different treatment of the norm operator determines the type of boson expansions as Hermitian or non-Hermitian.
The mapping operator limits the number of phonon excitations in addition to the phonon excitation modes beforehand to use the ideal boson state vectors as physical and avoid the breakdown of the small parameter expansion whose zeroth-order approximation is the boson approximation.

In the case that the closed algebraic approximation or phonon truncation approximation holds, that is, the double commutation relations between the phonons with excitation modes adopted as boson excitations are closed, the small parameter expansion is not available.
The norm operator is expressed as a function of the boson number operator,
which substantially makes all types of boson expansions be of finite expansion.

The small parameter expansion is not compatible with the closed-algebra approximation or the phonon truncation approximation.
The contribution of the phonon excitation modes neglected by the approximation makes the boson expansion become infinite expansion regardless of whether it is of the Hermitian type or not.
We have obtained the higher-order terms of the boson expansion not expanded so far in addition to the neglected by the approximation.

Conventional practical boson expansion methods have used the closed-algebra approximation or the phonon truncation approximation without recognizing its playing role mentioned above, and the claims derived from this approximation have no validity:
The normal-ordered linked-cluster expansion theory has failed to refute Marshalek's claim that KT-1 and KT-2 are of the chimerical boson expansion. The Dyson boson expansion theory does not have exceptional superiority over the other types of boson expansions.

The boson-fermion expansion theory derives the same boson expansions with the Hermitian-type boson expansions obtained here up to the next-to-leading order.
The boson-fermion expansion theory should derive higher-order expansion terms for further comparison.

Previous studies using the boson expansion methods should be re-examined by the norm operator method.

\appendix
\label{app:pair}
\section{formulae of the product of the pair operators}
We denote $B_q$, $X_{t''}$, or $X_{\bar t'}$ as $O_F$. the following equations hold:
\begin{subequations}
\begin{equation}
\widetilde{X_{t'}O_F}=\breve 1_B(X_{t'})_LB(O_F)_L\mathcal{\hat Z}\breve 1_B=\breve 1_B(X_{t'})_LB(O_F)_L\mathcal{\hat Z}\breve 1_B,
\end{equation}
\begin{equation}
\widetilde{O_F^\dagger X_t^\dagger}=\breve 1_B(O_F^\dagger)_L(X_t^\dagger)_L\mathcal{\hat Z}\breve 1_B.
\end{equation}
\end{subequations}

\begin{equation}
\widetilde{X_t^\dagger X_{t'}}=\breve 1_B\left\{(X_t^\dagger)_D(X_{t'})_D-\frac 12\sum_{t_1t_2}\sum_{\bar t'}Y(tt_1t_2\bar t'_1)b_{t_1}^\dagger b_{t_2}^\dagger (X_{t'})_L(X_{\bar t'_1})_L\right\}\mathcal{\hat Z}\breve 1_B.
\end{equation}

\begin{subequations}
\begin{equation}
\widetilde{X_{\bar t}^\dagger X_{\bar t'}^\dagger}=\breve 1_B(X_{\bar t}^\dagger)_L(X_{\bar t'}^\dagger)_L\mathcal{\hat Z}\breve 1_B+O(\Gamma^5),
\end{equation}

\begin{equation}
\widetilde{X_{\bar t} X_{\bar t'}}=\breve 1_B(X_{\bar t})_L(X_{\bar t'})_L\mathcal{\hat Z}\breve 1_B+O(\Gamma^5).
\end{equation}
\end{subequations}

\begin{equation}
\widetilde{X_{\bar t}^\dagger X_{\bar t'}}=\breve 1_B(X_{\bar t}^\dagger)_L(X_{\bar t'})_L\mathcal{\hat Z}\breve 1_B+O(\Gamma^5).
\end{equation}

\begin{subequations}
\begin{equation}
\widetilde{ B_qX_{\mu'}}=(B_q)_D\widetilde{X_{\mu'}}+\sum_{t}\sum_{\bar t'}\Gamma_q^{\bar t't}b_t^\dagger\widetilde{X_{\bar t'}X_{\mu'}},
\end{equation}
\begin{equation}
\widetilde{X_{\mu}^\dagger B_q}=\widetilde{X_{\mu}^\dagger}(B_q)_D+\sum_{t'}\sum_{\bar t}\Gamma_q^{t'\bar t}\widetilde{X_{\mu}^\dagger X_{\bar t}^\dagger}b_{t'}.
\end{equation}
\end{subequations}

\section{Proof of  $\mathbf{W}(N)=\mathbf{0}(N)$ for $N\ge 3$}
\label{app:pr}
$\hat Z(N)$ is related to $\hat Z^{(A)}(N)$ as
\begin{equation}
\hat Z(N)=\breve 1_B\hat Z^{(A)}(N)\breve 1_B=\hat 1_B(N)\hat Z^{(A)}(N)\hat 1_B(N).
\end{equation}
Introducing
\begin{equation}
\begin{array}{lll}
\hat Z'(N)&=&(\breve 1_B^{(A)}-\breve 1_B)\hat Z^{(A)}(N)(\breve 1_B^{(A)}-\breve 1_B)
\\
&=&(\hat 1_B^{(A)}(N)-\hat 1_B(N))\hat Z^{(A)}(N)(\hat 1_B^{(A)}(N)-\hat 1_B(N)),
\end{array}
\end{equation}
and
\begin{equation}
\hat W(N)=\breve 1_B\hat Z^{(A)}(\breve 1_B^{(A)}-\breve 1_B)
=\hat1_B(N)\hat Z^{(A)}(\hat 1_B^{(A)}(N)-\hat 1_B(N)),
\end{equation}
we obtain
\begin{equation}
\hat Z(N)^{(A)}=\hat Z(N)+\hat W(N)+\hat W(N)^\dagger +\hat Z'(N).
\end{equation}
Eq. (\ref{eq:zwwtz'}) expresses this relation as those of the matrices where $\mathbf{Z}^{(A)}(N)$, $\mathbf{Z}(N)$,
$\mathbf{Z' }(N)$, and $\mathbf{W}(N)$ are matrices representing $\hat Z^{(A)}(N)$, $\hat Z(N)$, $\hat Z'(N)$, and $\hat W(N)$, respectively.

If $Y(t'_1\bar t\mu t'_2)=0$, then $\hat W(2)=0$.
On the other hand,  from Eq. (\ref{eq:zn}), we obtain
\begin{equation}
\label{eq:zan}
\hat Z^{(A)}(N)=\frac 1N\sum_\mu(X_\mu^\dagger)_D\hat Z(N-1)^{(A)}b_\mu.
\end{equation}
If $\hat W(N-1)=0$, then
\begin{equation}
\label{eq:zanw0}
\hat Z^{(A)}(N)=\frac 1N\sum_\mu(X_\mu^\dagger)_DZ(N-1)b_\mu
+\frac 1N\sum_\mu(X_\mu^\dagger)_D\hat Z'(N-1)b_\mu.
\end{equation}
On the othrer hand, $\hat 1_B(N-1) b_\mu(\hat 1_B^{(A)}(N)-\hat 1_B(N))=0$ and $\hat 1_B(N)(X_\mu^\dagger)_D(\hat 1_B^{(A)}(N-1)-\hat 1_B(N-1))=0$ hold. Therefore if $\hat W(N-1)=0$, then $\hat W(N)=0$. That is $\hat W(N)=0$ for $N\ge 3$, and then $\mathbf{W}(N)=\mathbf{0}(N)$ for $N\ge 3$.
\end{document}